\documentclass[11pt,fleqn]{article}
\usepackage{amsmath,amsfonts,amssymb,cite}
\usepackage{graphicx}

\def\onecol{\onecolumn \mathindent 2em}

\def\noi{\noindent}

\newcommand{\Title}[1]{\noi {\Large\bf #1}\\[1ex]}
\newcommand{\Author}[2]{\noi{\bf #1}\\[2ex]\noi{\normalsize\it #2}\\}
\newcommand{\Abstract}[1]{\vskip 2mm \begin{center}
        \parbox{16.4cm}{\small\noi #1} \end{center}\medskip}

\def\nqq{\hspace*{-2em}}
\def\nhq{\hspace*{-0.5em}}

\def\cm{\hspace*{1cm}}

\def\al{&\nhq}
\def\lal{&&\nqq {}}
\def\eq{Eq.\,}
\def\eqs{Eqs.\,}
\def\beq{\begin{equation}}
\def\eeq{\end{equation}}
\def\bear{\begin{eqnarray}}
\def\bearr{\begin{eqnarray} \lal}
\def\ear{\end{eqnarray}}
\def\earn{\nonumber \end{eqnarray}}
\def\nn{\nonumber\\ {}}
\def\nnv{\nonumber\\[5pt] {}}
\def\nnn{\nonumber\\ \lal }
\def\nnnv{\nonumber\\[5pt] \lal }
\def\yy{\\[5pt] {}}
\def\yyy{\\[5pt] \lal }
\def\eql{\al =\al}

\def\then{\ \Rightarrow\ }
\def\d{\partial}
\def\e{{\rm \,e}}

\def\diag{\mathop{\rm diag}\nolimits}

\def\const{{\rm const}}
\def\eps{\varepsilon}

\def\half{{\tfrac 12}} 
\def\Half{{\dfrac 12}}
\def\eqn#1{\eq\eqref{#1}}
\def\rf{\eqref}

\def\th{_{\rm th}}
\def\mn{_{\mu\nu}}
\def\MN{^{\mu\nu}}
\def\mN{_\mu^\nu}

\def\M{{\mathbb M}}
\def\N{{\mathbb N}}
\def\R{{\mathbb R}}

\def\tT{{\tilde T}}
\def\sph{spherically symmetric}
\def\ssph{static, spherically symmetric}
\def\cy{cylindrical}
\def\cyl{cylindrically symmetric}
\def\bh{black hole}
\def\bhs{black holes}
\def\wh{wormhole}
\def\whs{wormholes}
\def\asflat{asymptotically flat} 
\def\emag{electromagnetic}
\def\Scw{Schwarz\-schild}
\def\RN{Reiss\-ner-Nord\-str\"om}
\def\Lem{Lema\^{\i}tre}

\usepackage{color}

\begin{document}
\onecol
\thispagestyle{empty}

\Title{Some unusual wormholes in general relativity}

\Author{K. A. Bronnikov\footnote{E-mail: kb20@yandex.ru}}
		{Center for Gravitation and Fundamental Metrology, VNIIMS, 
		Ozyornaya ul. 46, Moscow 119361, Russia;\\
		Institute of Gravitation and Cosmology, 
		Peoples' Friendship University of Russia (RUDN University),\\ 
		\cm	ul. Miklukho-Maklaya 6, Moscow 117198, Russia;\\
		National Research Nuclear University ``MEPhI'', 
		Kashirskoe sh. 31, Moscow 115409, Russia}


\Abstract
  {In this short review we present some recently obtained traversable wormhole
  models in the framework of general relativity (GR) in four and six dimensions
  that somehow widen our common ideas on \wh\ existence and properties. 
  These are, first, rotating cylindrical wormholes, \asflat\ in the radial direction and 
  existing without exotic matter. The topological censorship theorems are not violated 
  due to lack of asymptotic flatness in all spatial directions. Second, these are
  cosmological wormholes constructed on the basis of the \Lem-Tolman-Bondi solution. 
  They connect two copies of a closed Friedmann world filled 
  with dust, or two otherwise distant parts of the same Friedmann world.
  Third, these are wormholes obtained in six-dimensional GR, whose one entrance is
  located in ``our'' \asflat\ world with very small extra dimensions while the other ``end''
  belongs to a universe with large extra dimensions and therefore different physical 
  properties. The possible observable features of such \whs\ are briefly discussed.
}
\tableofcontents
\bigskip

\section{Introduction}

  Wormhole physics has become quite a popular research area, even though nobody has 
  ever seen a real \wh\ in Nature. Thus, on August 1, 2021, a search for the term ``\wh'' 
  on the site arxiv.org gave 2,154 results for all years and 303 results for past 12 months. 
  This popularity looks natural since a \wh\ is one more, in addition to a \bh, and even 
  simpler manifestation of a strongly curved space-time which can lead to many effects 
  of great interest.
    

  The term ``\wh'' is multi-valued: while it originally means a kind of tunnel or shortcut 
  between different space-times or between otherwise distant regions of the same space-time 
  (it is also called a Lorentzian \wh), many authors discuss what they call quantum and 
  Euclidean \whs, giving this name to wave functions or objects in spaces with Euclidean
  signature whose properties resemble those of ``tunnels''. They are not our subject here. 
  Among Lorentzian \whs, one often speaks of traversable or nontraversable 
  ones, and some of them can also be one-way traversable. A lack of two-way 
  traversability is generally caused by the existence of horizons, and, in my opinion, \
  it is then more correct to call such objects \bhs\ thus avoiding their confusion with \whs. 
  Curiously, the first mentions of wormhole-like geometries (though, purely spatial ones) 
  by Flamm \cite{flamm} and by Einstein and Rosen \cite{ER} actually 
  referred to the \Scw\ and \RN\ \bh\ space-times. In this paper we are going to discuss 
  only two-way traversable \whs\ without indicating this each time.
  
  Another issue is that of ``unusual'' \whs. Which of them could be called ``usual'' if none have 
  been observed as yet? The answer is --- those which are most frequently discussed, hence, 
  spherically or axially symmetric ones, possibly with rotation, connecting spatial regions 
  with similar properties, those which are most frequently \asflat\ or asymptotically AdS and, 
  if considered in the framework of GR, necessarily require for their support some kind of 
  exotic (or phantom) matter, violating the Null Energy Condition (NEC), a part of the 
  Weak Energy Condition (WEC) that provides nonnegativity of the energy density in any 
  reference frame. In GR extensions, the role of exotic matter may be implemented by some 
  geometric quantities like torsion, nonmetricity or nonlinear curvature constructs, maybe of 
  extra-dimensional origin. There are many interesting tasks and problems on this trend, 
  connected with their possible observable features, their stability, evolution, etc., see, e.g., 
  the reviews \cite{rev1, rev2, BR21}, but we will leave them aside in the present paper. 
   
  One kind of ``unusual'' wormholes to be discussed here is connected with topological 
  features that make impossible asymptotic flatness in all spatial directions: these are 
  \cyl\ configurations, looking like cosmic strings when observed from afar. It turns out 
  that this structure allows one to avoid the well-known topological censorship theorems 
  \cite{censor1,censor2} and to construct stationary \wh\ models in GR without WEC 
  violation \cite{cyl1,cyl2,cyl3}. 
  
  Other \whs\ of interest are those cosmological in nature: they exist in the framework of
  cosmological models and may be, above all, relevant to early Universe descriptions,
  for example, those with two de Sitter asymptotic regions may connect otherwise distant 
  parts of an inflationary Universe \cite{wh-dS}, thus increasing its causal connection.
  We, however, focus here on the recent and somewhat unexpected inference 
  \cite{sushkov20, we21} that cosmological \whs\ may be supported by evolving 
  dust distributions and obtained based on the \Lem-Tolman-Bondi (LTB) famous solution 
  \cite{lemaitre, tolman, bondi}.
  
  Lastly, among the diverse \wh\ models in multidimensional gravity theories, we will 
  single out and discuss a curious class of \whs\ in six-dimensional (6D) GR 
  \cite{ex-D1, ex-D2} able to connect our 4D space-time possessing small and 
  invisible extra dimensions with another one, where the extra dimensions are large, 
  or maybe with a multidimensional region of our own Universe if such regions do exist.
  
  Sections 2--4 are devoted to these three kinds of unusual \whs, and Section 5 
  is a conclusion.  
    
\section{Cylindrical symmetry and phantom-free models}

  An important topological censorship theorem \cite{censor1} asserts that in GR,
  if an asymptotically flat, globally hyperbolic space-time ($\M, g\mn$) satisfies the
  averaged NEC, then every causal curve from past null infinity to future null infinity
  is homotopic to a topologically trivial curve of this kind. As a consequence, if there 
  emerges a \wh-like structure, it cannot be traversed even by light, to say nothing 
  on massive bodies, probably due to a rapid collapse of such an object. Somewhat 
  weaker versions of this theorem concern static \whs, proving the necessity of NEC 
  violation at \wh\ throats, i.e., their narrowest parts \cite{thorne, hoh-vis}.
  
  Is it possible to circumvent these theorems? A straightforward way of obtaining 
  \wh\ solutions in GR is to violate the NEC by introducing some kind of exotic or 
  phantom matter, for example, a scalar field with negative kinetic energy or thin shells,
  as is done in a lot of studies. One can also recall such axially symmetric wormhole
  structures in GR as the Kerr metric with a supercritical angular momentum, Zipoy's 
  static solutions \cite{zipoy}, their extensions with electromagnetic and scalar fields 
  \cite{ax1, ax2, ax3}: each of them connects two different flat infinities, being free from 
  exotic matter but, instead, containing a naked ring singularity around a disk-shaped 
  throat, thus violating global hyperbolicity.
  
  One more way to phantom-free \whs\ is to replace GR with some extended theory
  of gravity, delegating the role of exotic matter to geometric quantities like torsion 
  or the Gauss-Bonnet invariant, see the reviews \cite{rev1, rev2, BR21}. 
  
  Let us. however, discuss the fourth way: remaining in the framework of GR,   we 
  abandon asymptotic flatness in all directions, assuming, in the simplest case, \cy\ symmetry. 
  To describe a stringlike object observable from a weakly curved region of space, a 
  \cy\ metric should be \asflat\ in the radial directions, whereas in the longitudinal direction
  the curvature does not change and remains arbitrarily large.
  
\subsection{Definitions of a cylindrical \wh}  

  A stationary \cyl\ space-time can be described by the metric 
\beq                      \label{ds-cy}
		ds^2 = \e^{2\gamma(x)}[ dt - E(x)\e^{-2\gamma(x)}\, d\varphi ]^2
       - \e^{2\alpha(x)}dx^2 
	- \e^{2\mu(x)}dz^2 - \e^{2\beta(x)}d\varphi^2,
\eeq  
  where $x$, $z\in \R$ and $\varphi\in [0, 2\pi)$ are the radial, longitudinal and 
  angular coordinates, respectively, and $x$ is specified up to a substitution
  $x \to f(x)$, hence its range depends both on its choice (``gauge'') and on the 
  geometry under consideration. The quantity $r(x) = \e^\beta$ is the circular radius,
  whose possible regular minimum can be identified as a \wh\ throat (a so-called ``r-throat'').

  There is an alternative definition of a throat (an ``a-throat'') as a regular minimum of the 
  ``area function'' $a(x) = \e^{\beta + \mu}$, but it is in general less applied since 
  $r(x)$ is more evident. Also, if the whole configuration is \asflat\ at large radii 
  on both ends of the $x$ range, it evidently means that there are throats according 
  to both definitions. Therefore we will pay attention to both of them.
  
  The off-diagonal metric component $E$ describes rotation, unless 
  $E\e^{-2\gamma} = \const$;
  in the latter case the metric is static, and $E=0$ is easily obtained by redefining the 
  time coordinate ($ t \mapsto t + \const\cdot\varphi$). 
  
  A \wh\ with the metric \rf{ds-cy} may be defined as a regular space-time region 
  with finite functions $\alpha, \beta, \gamma, \mu, E$, containing a throat $x = x\th$
  (by one or both of the above definitions), and such that the radius $r(x)$ 
  and/or the area function $a(x)$ reach values much larger than at $x=x\th$ on both 
  sides of the throat.
  
\subsection{A static \wh\ with a magnetic field source}

  It is not a problem to find solutions to the Einstein equations with r-throats and non-phantom 
  sources. One such example is a static solution ($E=0$) with an azimuthal magnetic field 
  \cite{cyl-EM, cyl-wh, cyl-rev} written in terms of the harmonic radial coordinate $x$
  satisfying the gauge condition $\alpha = \beta+\gamma+\mu$:
\bearr           \label{ds-az}
     	 ds^2 =  \frac{q^2 \cosh^2 kx}{k^2}
     	 		\Big[e^{2a} dt^2 - e^{2(a+b)x} dx^2 - e^{2bx} d\phi^2 \Big] 
     	 				- \frac{k^2 dz^2}{q^2 \cosh^2 kx},
\nnn
		q, a, b, k = \const >0; \qquad  ab = k^2;
\nnn		 
	F^{xz} = q\e^{-2\alpha(x)}	, \qquad F_{xz} = q\e^{2\mu(z)};  
	\qquad  B^2 = q^2 \e^{2\mu - 2\alpha} = \frac{k^4 \e^{-2(a+b)x}}{q^2 \cosh^4 kx},
\ear
  where $x \in \R$, $F\mn$ is the electromagnetic field tensor, and $B$ is the 
  $\varphi$-directed magnetic induction created by an effective current $q$ 
  along the $z$ axis, The solution \rf{ds-az} describes a \wh\ if $a > k$ since in 
  this case $r = \e^\beta \to \infty$ at both ends, $x \to \pm \infty$. The \wh\ is highly 
  asymmetric, for example, $B \to 0$ at the right end ($x\to \infty$) and to infinity
  at the left end, from which it is clear that $x \to -\infty$ is a singularity.
  Meanwhile, we have $\e^\mu \to 0$ (a shrinking longitudinal scale) on both ends.
  Moreover, the area function $a(x) = \e^{\beta+\mu} = \e^{bx}$ has no minimum, and
  \rf{ds-az} is a \wh\ solution only by the definition related to $r(x)$.

\subsection{A rotating \wh\ with a scalar field source}
  
  Another example is a rotating \cy\ \wh\ with a massless scalar field $\phi$ \cite{cyl-rot}
  having the Lagrangian $L_\phi = \half \eps \phi^{,\mu} \phi_{,\mu}$, where $\eps = +1$ 
  corresponds to a canonical scalar field, and $\eps = -1$ to a phantom one. 
  In terms of the metric \rf{ds-cy} and the harmonic radial coordinate $x$, the solution reads
\bearr                     \label{sol-phi}
		\e^{2\alpha} = \e^{(4m - 2h)x}, \qquad  \qquad
		\e^{2\beta} = \frac{k\e^{2hx}}{2\omega_0 \sin kx},
\nnn
		\e^{2\gamma} = \frac{2\omega_0}{k} \e^{2hx} \sin kx, \qquad 
		\e^{2\mu} = \e^{-2mx},  \qquad    \phi = Cx,			
\nnn
		E = \frac{\e^{2hx}}{k}\big[E_0 \sin kx + k \cos kx\big], \qquad
		\omega = \omega_0 \e^{-\mu - 2\gamma} = \frac {k\e^{(m-2h)x}}{2 \sin kx},	
\ear
  where $\omega$ is the vorticity defined as the angular velocity of tetrad rotation
  \cite{krechet}, while $C, h, k, m, E_0$ and $\omega_0$ are integration constants related by 
\beq 				\label{sol-phi1}
 		k^2 = 4h(2m-h) - 2\kappa \eps C^2,
\eeq   
   where $\kappa = 8\pi G$ is the gravitational constant.

  The range of the coordinate $x$ in \rf{sol-phi} is $x \in (0, \pi)$, and its both extremes 
  possess $r(x) = \e^\beta \to \infty$ and $a(x) = \e^{\beta+\mu} \to \infty$, but the metric is
  singular there due to $g_{tt} = \e^{2\gamma} \to 0$. It can also be noticed that this 
  \wh\ solution exists with both phantom and canonical scalar fields ($\eps = \pm 1$) 
  and even in the vacuum case ($C =0 \then \phi =0$). 
  
  It is known, however, that to support a static \cy\ \wh\ with an a-throat, its matter source 
  should necessarily violate the WEC and NEC \cite{cyl-wh}. Hence it is rotation that provides
  the \wh\ nature of \rf{sol-phi}.
  
\subsection{The rotational part of Ricci and Einstein tensors}

  To explain why rotation favours the emergence of \whs, let us assume that the matter 
  source of the metric \rf{ds-cy} is taken in its comoving reference frame, in particular, its 
  velocity in the $\varphi$ direction is zero. Then we can directly integrate the Einstein equation 
  $R^3_0 \propto (\omega \e^{2\gamma+\mu})' =0$ (the prime stands for $d/dx$) to
  obtain $\omega = \omega_0 \e^{-\mu -2\gamma}$ with $\omega_0 = \const$. Then 
  the nonzero components of the Ricci tensor for the metric \rf{ds-cy} have the form
\bear                 \label{Ric}
      R^0_0 \eql -\e^{-2\alpha}[\gamma'' + \gamma'(\sigma' -\alpha')] - 2\omega^2,
\nnv      
      R^1_1 \eql -\e^{-2\alpha}[\sigma'' + \sigma'{}^2 - 2U - \alpha'\sigma']+ 2\omega^2,
\nnv      
      R^2_2 \eql -\e^{-2\alpha}[\mu'' + \mu'(\sigma' -\alpha')],  
\nnv      
      R^3_3 \eql -\e^{-2\alpha}[\beta'' + \beta'(\sigma' -\alpha')] + 2\omega^2, 
\nnv
      R^0_3 \eql G^0_3 =  E \e^{-2\gamma}(R^3_3 - R^0_0), 
\ear
  under an arbitrary choice of the coordinate $x$, with the notations
\beq
            \sigma = \beta + \gamma + \mu, \qquad
            U = \beta'\gamma'  + \beta'\mu' + \gamma' \mu'.
\eeq     
  We see that the rotational parts of the Ricci tensor and the Einstein tensor 
  $G\mN = R\mN - \half \delta\mN R$ are separated from their static parts, 
  corresponding to \rf{ds-cy} with $E\equiv 0$:
\bearr                      \label{vortex}
		R\mN = R\mN{}_{\rm (st)} + R\mN{}_{\rm (rot)},\qquad
             			          R\mN{}_{\rm (rot)} = \omega^2 \diag(-2, 2, 0, 2),
\nnn
          G\mN = G\mN{}_{\rm (st)} + G\mN{}_{\rm (rot)},\qquad
             			          G\mN{}_{\rm (rot)} = \omega^2 \diag(-3, 1, -1, 1).
\ear
  In the Einstein equations $G\mN = - \kappa T\mN$, the stress-energy tensor (SET) 
  $T\mN$ thus actually acquires quite an exotic contribution $G\mN{}_{\rm rot}/\kappa$ 
  due to rotation, with negative effective energy density equal to $-3 \omega^2/\kappa$.
  
\subsection{How to obtain \asflat\ rotating \whs} 

  If our aim is to obtain phantom-free \cy\ \whs\ which could be observable from weakly
  curved regions of space on each side from the throat, we should require their asymptotic
  flatness in the radial direction at both ends of the $x$ range. We have seen that it
  is impossible with static \whs\ due to conditions on an a-throat. Rotation allows for
  more or less easily finding solutions with both kinds of throats, but it becomes quite a 
  hard problem to obtain asymptotic flatness. The following trick was suggested in 
  \cite{cyl-rot} to overcome this difficulty: having obtained a phantom-free \cy\ \wh\ 
  solution, cut out from it a regular region containing the throat and join it to two flat regions 
  extended to infinity on each side from the throat. The whole system will thus become 
  twice asymptotically flat and potentially observable from each side. To form a single 
  space-time in this way, the external and internal metrics on the two junction surfaces 
  $\Sigma_+$ and $\Sigma_-$ should be identified, but the derivatives of the metric 
  in the direction across $\Sigma_+$ and $\Sigma_-$ --- that is, in the radial direction ---
  will in general suffer discontinuities, which, according to the Darmois-Israel formalism 
  \cite{darmois, israel}, mean that there are some surface matter distributions on 
  $\Sigma_\pm$. Then, to obtain a completely phantom-free \wh\ model, we should 
  require that this surface matter should satisfy the WEC (and NEC as its part).  
  
  The first attempts to realize this program were undertaken in \cite{cyl-rot} using 
  rotating \wh\ solutions with scalar fields with or without self-interaction potentials.
  With such interior solutions, it turned out to be impossible to satisfy the WEC on
  both surfaces $\Sigma_pm$. Later on, a no-go theorem was proved \cite{phi-nogo},
  claiming that the threefold construction described above cannot be phantom-free 
  if the SET of matter in the internal (\wh) region satisfies the equality 
  $T^t_t =T^z_z$. This equality holds for all kinds of scalar fields $\phi(x)$ minimally
  coupled to gravity, for example, those of generalized k-essence type, with an 
  arbitrary Lagrangian of the form $L(\phi, X)$ with $X = g\MN \phi_{,\mu}\phi_{,\nu}$. 
  
  Successful models of this kind were obtained later in \cite{cyl1} with a \wh\ solution
  sourced by a certain kind of anisotropic fluid, in \cite{cyl2} with an isotropic maximally 
  stiff perfect fluid with the equation of state $p=\rho$, and in \cite{cyl3} with a source 
  consisting of a perfect fluid with $p = w\rho$ ($|w| < 1$), and a magnetic field along 
  the $z$ axis whose source is a circular electric current in the angular direction.
  The model of \cite{cyl3} tends to that of \cite{cyl2} in the limit $w\to 1$, while in the 
  same limit the azimuthal current and the magnetic field vanish. In the next subsection 
  we briefly describe the model of \cite{cyl2}.

\subsection{An \asflat\ rotating \wh\ model with stiff matter} 

  In a perfect fluid with the equation of state $p=\rho$, the speed of sound is equal to
  the speed of light, which greatly simplifies the equations of motion and has led to 
  obtaining numerous solutions which are either unique or much simpler than their 
  counterparts with other equations of state: see, e.g., \cite{exact-book, cyl-rev} for a 
  general exposition,  \cite{safko72, evans77, cyl-flu1, cyl-flu2} for static solutions, 
  \cite{stiff-w} for wavy ones, and \cite{santos82, sklav99, ivanov02, santos06, aniso} for 
  stationary ones; applications of stiff matter in modern theoretical cosmology can also
  be mentioned, see, e.g., \cite{stiff1, stiff2}; see also references therein. 

  As before, let us use the Einstein equations for the metric \rf{ds-cy}
  with the harmonic radial coordinate, $\alpha = \beta+\gamma+\mu$. 
  For matter density, the conservation law gives $\rho = \rho_0 \e^{-2\gamma}$.
  Furthermore, one of the equations reads simply 
\beq   
	\mu'' =0  \ \then \        \mu = mx + \mu_0, \cm m, \mu_0=\const,
\eeq  
   and we put $\mu_0=0$ by properly rescaling the $z$ axis. Other equations then imply
\beq     
                   \beta'' = 2\omega_0^2 \e^{2\beta-2\gamma}, \cm
                   \beta'' + \gamma'' = 2\kappa\rho_0 \e^{2\beta + 2mx}.  
\eeq          
  For simplicity let us assume symmetry under reflections $x \to -x$, so that $m=0$,
  and $\gamma \equiv 0$. We then obtain
\beq                \label{be''}
		  \beta'' = 2\omega_0^2 \e^{2\beta},    \cm \kappa\rho_0 = \omega_0^2.
\eeq      
  From the three branches of solutions to the Liouville equation for $\beta$, we choose 
  the one where the function $\beta(x)$ has a minimum, necessary for obtaining a \wh\ 
  solution. We thus have
\beq
		\e^{\beta} = \frac {k}{\sqrt{2\omega_0^2}\cos (kx)}, \cm k = \const> 0,
		\cm x \in (-\pi/2, \pi/2) 
\eeq      
  under a proper choice of the zero point of $x$. It remains to find $E(x)$, 
  and to do that we use the expression following from the definition of $\omega(x)$ 
  in terms of $E$ and the expression of $\omega(x)$ implied by the assumed 
  comoving nature of the reference frame (see, e.g, \cite{cyl-rot, cyl2}):
\beq
 		\omega(x) = \Half(E\e^{-2\gamma})' \e^{\gamma-\beta-\alpha}
 				= \omega_0 \e^{-\mu - 2\gamma}
\eeq  
  As a result, we have   
\beq           \label{E1}
		E = 2\omega_0 \e^{2\gamma} \int \e^{\alpha+\beta-\mu-3\gamma} dx 
		= \frac{1}{\omega_0} \int \frac{k^2 dx}{\cos^2(kx)} = \frac{k}{\omega_0} \tan (kx),
\eeq
  where we have chosen the integration constant so that $E(x)$ is an odd function,
  it is convenient for the subsequent matching with the exterior Minkowski metric. 
  For the same purpose we introduce an arbitrary time scale by putting 
  $dt \mapsto \sqrt{P} d\tau$, $P = \const > 0$. We also have $\kappa \rho_0 = \omega_0^2$.
  
  Thus we know the solution completely. It is most conveniently expressed in terms of 
  a new radial coordinate, $y = k \tan(kx)$:
\beq                      \label{x-y}
             dx = \frac{dy}{k^2 + y^2}, \qquad	\e^{2\beta} = \frac{k^2+y^2}{2\omega_0^2 }, 
             \qquad  E = \frac{\sqrt{P} y}{\omega_0},
\eeq
  and consequently
\beq                                       \label{ds-wh}  
  	   ds^2 = \bigg(\sqrt{P} dt -  \frac{y}{\omega_0} d\varphi\bigg)^2 
  	   - \frac{dy^2}{2\omega_0^2(k^2+y^2)} - dz^2 - (k^2 + y^2) \frac{d\varphi^2}{2\omega_0^2}.
\eeq  
  The solution is regular at all $y\in \R$, and a minimum of $r(x)$ occurs at $y =0$. 
  One can notice, however, that at $y^2 > k^2$ it becomes $g_{33} > 0$, so that the 
  coordinate circles $0 < \phi < 2\pi$ are closed timelike curves, violating causality.
  
  Having obtained this \wh\ solution, let us try to match it at some values of $y$, both positive 
  and negative, to the Minkowski metric, taken in a rotating reference frame:     
\beq                                                          \label{ds_M}
	      ds_{\rm M}^2 = dt^2 - dX^2 - dz^2 - X^2 (d\varphi + \Omega dt)^2,
\eeq
  so that in the notations of (\ref{ds-cy}) we have
\bearr                                                   \label{M-param}
      \e^\alpha = 1, \qquad  \e^{2\gamma} =  1 - \Omega^2 X^2,\qquad
      \e^{2\beta} = \frac{X^2}{1 - \Omega^2 X^2},\qquad
      E = \Omega X^2, \qquad   \omega = \frac{\Omega}{1 - \Omega^2 X^2},
\nnn 
      \gamma' = -\frac{\Omega X}{1 - \Omega^2 X^2}, \qquad   \mu' =0, \qquad
      \beta'  = \frac 1X  + \frac{\Omega^2 X}{1 - \Omega^2 X^2}.
\ear  
  
   If we identify a  surface $y=y_0$ in the \wh\ space-time \rf{ds-wh} 
   with a surface $X=X_0$ in \rf{ds_M}, the matching conditions are
\beq                                                       \label{ju-1}
      [\beta] = 0, \quad [\mu] = 0, \quad     [\gamma] = 0, \quad [E] =0,
\eeq  
  where, as usual, for any $f(x)$, $[f]$ denotes its jump at the junction $\Sigma$. 
  With \rf{ju-1}, the coordinates $t, z, \phi$ can be identified in the whole space.
  The choice of the radial coordinates may be different on different sides from 
  $\Sigma$, but it is unimportant since all quantities involved in the matching conditions 
  are insensitive to this coordinate choice.
  
  The junction surface $\Sigma$ having been identified, we can determine 
  its material content using the Darmois-Israel formalism  \cite{darmois,israel,berezin87}: 
  its SET $S_a^b$ is found in terms of the extrinsic curvature $K_a^b$ as
\bearr                                                        \label{ju-2}
	S_a^b =  \kappa^{-1} [\tilde K_a^b], \qquad
                  		\tilde K_a^b := K_a^b - \delta_a^b K^c_c, 		
\ear
  where $a, b, c = 0, 2, 3$. The WEC requirements for the surface SET read
\beq                                                             \label{WEC}
	S_{00}/g_{00} = \sigma_s \geq 0, \qquad\ S_{ab}\xi^a \xi^b \geq 0,
\eeq
  where $\sigma_s$ is the surface energy density and $\xi^a$ an arbitrary null
  vector on $\Sigma$. As shown in \cite{cyl1, cyl2}, the requirements \rf{WEC} reduce to
\bear       \label{Wa} 
	a+c+\sqrt{(a-c)^2 +4 d^2} &\geq & 0,
\\          \label{Wb}    
	a+c+\sqrt{(a-c)^2 +4 d^2} + 2b & \geq & 0,
\\            \label{Wc}
	a + c & \geq & 0,
\ear       
   with
\beq     
           a = \big[\e^{-\alpha}(\beta'+\mu')\big], \qquad
           b = \big[\e^{-\alpha}(\beta'+\gamma')\big], \qquad
           c = \big[\e^{-\alpha}(\gamma' + \mu')\big], \qquad   d = -[\omega],
\eeq
  where the prime denotes a derivatives in the radial coordinate used in the corresponding 
  spatial region. 
   
  Let us apply these conditions, beginning, for certainty, with the surface $\Sigma_+$ 
  ($y=y_0 >0$, $X=X_0 > 0$). The condition $[\mu] =0$ is trivial, while the other three 
  conditions \rf{ju-1} give
\beq                \label{match1}
                    P = 1 - \Omega^2 X^2, \cm
                    \Omega X^2 = \frac{y\sqrt{P}}{\omega_0}, \cm 
                    \frac{k^2+y^2}{2\omega_0^2} = \frac{X^2}{P}.
\eeq       
  where the index ``zero'' at $X$ and $y$ is omitted with no risk of confusion, and we
  assume $\omega_0 > 0$. There are initially six independent parameters
  [$P,  k, \omega_0, y$ in the internal  metric \rf{ds-wh} and $\Omega, X$ in \rf{ds_M}], 
  and they are connected by three conditions \rf{match1}. We can choose 
  the following independent parameters: $X = X_0$ with the dimension of length
  (specifying the length scale of the model), and the dimensionless 
  $y =y_0$ and $P$, then for the remaining parameters we obtain:
\beq            \label{param1}  
		  \Omega = \frac{\sqrt{1-P}}{X} , \cm 
		  \omega_0 = \frac{\sqrt{P} y}{\sqrt{1-P}X}, \cm   
		  k^2 = y^2 \frac{1+P}{1-P},
\eeq        
  which leads to the following expressions for the quantities $a,b,c,d$ from \rf{Wa}:
\bearr                \label{abcd1}
	a = \frac{P^{3/2}y  -1}{PX}, \qquad
	b = \frac{1 - y\sqrt{P}}{X}, \qquad
	c = \frac{P-1}{PX}, \qquad
	d = \frac{\sqrt{P}y - 1 + P}{XP\sqrt{1-P}}.  
\ear      
 Ignoring the common factor $1/X$, it is straightforward to verify that  
  the requirements \rf{Wa}, \rf{Wb}, \rf{Wc} are fulfilled under the condition 
\beq            \label{y1} 
		y \geq \frac{2 - P}{P^{3/2}}.
\eeq    
  
  Now let us discuss the same requirements on $\Sigma_-$, specified by $X = -X_0 < 0$ 
  and $y= -y_0 <0$. In all expressions  \rf{abcd1} the factor $1/X$ is now negative. But, 
  at the same time, the signs of all discontinuities also change: on $\Sigma_+$ we took
  $[f] = f_{\rm out} - f_{\rm in}$ for every quantity $f$, whereas on $\Sigma_-$ the 
  opposite must be taken. As a result, the parameters $a,b,c$ preserve their form \rf{abcd1} 
  with $X$ replaced by $|X|$ (where we denote $y = |y_0| >0$).
  For $d = -[\omega]$ we must bear in mind that, by \rf{match1},
  $\Omega (\Sigma_-) = -\Omega (\Sigma_+)$, whereas in the internal solution 
  $\omega (\Sigma_-) = \omega (\Sigma_+)$, therefore, on $\Sigma_-$
\[
	d \ \mapsto \ d_- = - \frac {1 - P + |y| \sqrt{P}}{|X| P \sqrt{1-P}},
\]    
  so that $|d_-| > |d|$, making it even easier to fulfill the requirement \rf{Wb}. 
  Thus all WEC requirements are fulfilled under the same condition \rf{y1},
  and so this condition leads to a completely phantom-free \wh\ model.
  
  Moreover, from \rf{param1} it follows $y_0^2 < k^2$, hence $y^2 < k^2$
  in the whole internal region, therefore closed timelike curves are absent.
  For $P \in (0,1)$, according to \rf{y1}, the parameters $y=y_0$ and $k$ should
  be large enough, see Fig.\,1.
  
\begin{figure}[!h]
\centering
\includegraphics[width=45mm]{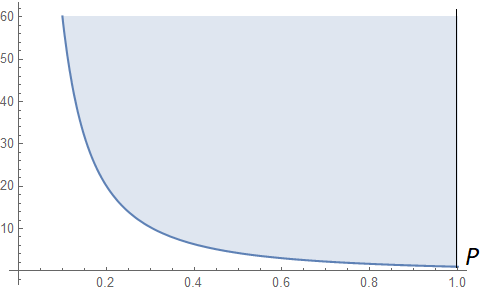}\quad
\includegraphics[width=45mm]{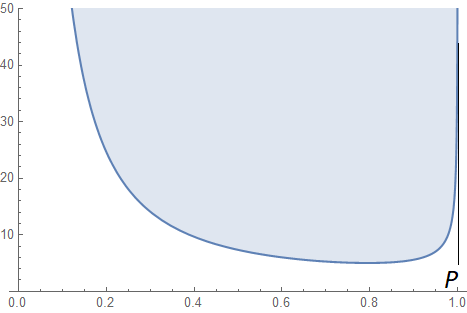}
\caption{\small 
	Admissible values of the parameters $y_0$ (left panel)
	and $k$ (right panel) in our \wh\ model are shown by gray areas.}
\label{fig1}
\end{figure}   

  The model described, though quite simple, looks rather artificial, but the very existence 
  of such models proves an important statement: it is possible to obtain phantom-free 
  \whs\ in the framework of GR, but at the expense of lacking asymptotic flatness in
  all directions. Other models of this kind have been constructed in \cite{cyl1, cyl3}. 
  
  To describe the external regions of \cy\ \whs, we have been using the Minkowski 
  metric in the form \rf{ds_M}. In the same way, we can consider, instead, a cosmic 
  string metric: to do so, it is sufficient to replace $X$ in all equations with 
  $X \sqrt{1 - {\tilde \alpha}/(2\pi)}$, where the angular deficit $\tilde \alpha$ 
  is interpreted as resulting from an effective linear density $\tilde\mu$ of the cosmic string 
  \cite{vilenk}, $\tilde \alpha = 16\pi^2 G \tilde\mu$. 
  
  In the case of flat asymptotic regions, a \wh\ like those considered here might probably
  be observed only when particles or celestial bodies reach and encounter its junction
  surface. In the case of the string asymptotic behavior, all observational properties 
  of cosmic strings should be added. 
  
  Two \ssph\ \wh\ models of interest have been recently constructed with sources 
  in the form of a pair of classical electrically charged spinor fields \cite{radu21, kono21},
  and they were claimed to be free from exotic matter. The latter statement evidently
  contradicts the topological censorship theorems, therefore it seems more correct to 
  say that the spinor fields in these models acquire exotic properties and violate
  the NEC \cite{comment}. A \ssph\ \wh\ in GR, supported by a three-form field with a 
  correct sign of kinetic energy, was described in \cite{bouh21}; the necessary NEC 
  violation is provided in this model by a suitable quartic potential.  

\section{Cosmological \whs}

\subsection{General observations}

  There are quite a lot of papers that connect the \wh\ concept with cosmology.
  This means, above all, that the asymptotic behavior cannot be flat and should 
  correspond to an expanding Universe. Wormholes themselves are in this case 
  mostly dynamic, for which the key notion of a throat can be introduced in different 
  ways, which in turn leads to different inferences on the \wh\ existence and 
  properties. The definitions by Hochberg and Visser \cite{hoch98} and Hayward
  \cite{hay99} use, though in different ways, the behavior of null congruences 
  and have been shown to require WEC violation from the supporting matter. 
  The definition suggested by Tomikawa, Izumi and Shiromizu \cite{tomikawa15}
  also rests on the properties of null congruences, but can avoid NEC violation in
  space-times with singularities \cite{bittencourt17}. Maeda, Harada and Carr
  \cite{maeda09} expressed an opinion that the definitions of \cite{hoch98, hay99} 
  are not well motivated in a cosmological background and suggested a more 
  intuitively clear definition of throats as surfaces of minimum area on spacelike
  hypersurfaces. Bittencourt, Klippert and Santos \cite{bittencourt17} compared 
  all these definitions in application to a particular \wh\ model and, for some reasons, 
  gave preference to Hayward's definition. We will here still prefer the one from
  \cite{maeda09} which directly generalizes the ``static'' definition and is more 
  frequently used by the researchers, despite its ambiguity due to arbitrariness 
  in choosing the appropriate family of spacelike hypersurfaces.
  
  With any definition of a throat, a \wh\ is understood as a space-time region 
  containing a throat and extending sufficiently far from this throat in both sides. 
  
  In this section we restrict ourselves to \sph\ space-times, both for simplicity and 
  because the great majority of dynamic \wh\ solutions obtained thus far are \sph.
 
  One class of what may be called cosmological \whs\ is represented by 
  formally static solutions to the Einstein equations with a cosmological constant 
  $\Lambda$: those with $\Lambda < 0$ are, in general, asymptotically 
  anti-de Sitter (AdS) (assuming that other matter sources decay rapidly 
  enough at infinity) and are globally static, while those with $\Lambda > 0$ are 
  asymptotically de Sitter and inevitably have cosmological horizons. So
  they are really cosmological and can connect otherwise distant parts of an
  expanding de Sitter universe or different de Sitter universes. 
  
  It was repeatedly noticed that the possible phantom nature of Dark Energy 
  responsible for the accelerated expansion of our Universe may be favorable 
  for \wh\ existence. This Dark Energy should possess isotropic pressure 
  to govern the expansion of our isotropic (at large) Universe. But a simple no-go
  theorem proved in \cite{wh-dS} shows that isotropic matter of any kind, be it
  phantom or not, cannot support \whs\ with flat and AdS asymptotics, whereas 
  the de Sitter behavior far from the throat is quite possible, and a number of 
  such solutions have been obtained and studied
  \cite{wh-dS, lemos03,tomikawa15}, see also references therein.
  
  Wormhole-like models somewhat similar to those with $\Lambda >0$
  but with a de Sitter asymptotic only at one end and flat at the other, or
  de Sitter at both ends but with different effective cosmological constants, can
  be obtained in models with phantom (or partly phantom, so-called ``trapped ghost'')
  scalar fields \cite{pha1, pha2, pha-em, pha-sca}.  
  
  Many other cosmological \wh\ models are constructed by multiplying a static 
  \wh\ metric by a time-dependent conformal factor which may be interpreted as
  a cosmological scale factor, e.g., \cite{kar94, sw_kim96, anchor97, sushkov07},
  and it was noticed that NEC violation can be avoided owing to this scale factor.
  Some models have been obtained by assuming special kinds of matter 
  as a source of gravity in time-dependent \whs\ with cosmological asymptotic 
  regions: a Chaplygin gas \cite{mokeeva12} and nonlinear electromagnetic
  fields with Lagrangians of the form $L(F)$, $F= F\mn F\MN$
  \cite{arellano06, NED18}. It has turned out that such \wh\ solutions require 
  some special forms of the function $L(F)$ with non-Maxwell behavior at small 
  $F$ \cite{NED18}. 

  Of special interest are attempts to obtain \whs\ with such a realistic source 
  as dust (a perfect fluid with zero pressure) due to their possible relation 
  to the matter-dominated stage of expansion in our Universe
  \cite{fara10, sushkov20}. With spherical symmetry, one then naturally uses
  the famous \Lem-Tolman-Bondi (LTB) solution to the Einstein equations
  \cite{lemaitre,tolman,bondi} or its generalization with radial electric or
  magnetic fields \cite{markov, bailyn, vickers, lapch, khlest, shik}.
  In \cite{fara10}, a \wh\ joining two spatially flat LTB universes was sustained 
  by a thin shell satisfying the WEC, which is an interesting but somewhat undesired
  feature. In \cite{sushkov20}, by properly choosing the arbitrary functions of the 
  LTB solution, the authors obtained a model of a collapsing dust ball with inner
  wormhole structure but globally forming a \bh. Our recent study \cite{we21},
  has confirmed that the LTB solution or its electric/magnetic generalization cannot 
  lead to an \asflat\ (even dynamic) \wh, but it is shown to be possible to obtain a 
  cosmological \wh\ connecting two closed isotropic universes or distant parts 
  of the same universe without any thin shells. The following subsections briefly 
  describe such a model.  

\subsection{LTB solution with a radial magnetic field}

  Let us begin with a brief presentation of the LTB solution generalized to include 
  a contribution from an external radial electric or magnetic field (to be called a q-LTB
  solution). In a comoving reference frame for neutral dust particles, the metric may be 
  written in the synchronous diagonal form
\beq                                                    \label{ds-tol}
      ds^2 = d\tau^2 - \e^{2\lambda(R,\tau)} dR^2 - r^2(R,\tau) d\Omega^2,
\eeq
  with $\tau$ the proper time along particle trajectories, and $R$ a radial coordinate
  specified up to a reparametrization $R \to F(R)$, and $d\Omega^2$ is the
  metric on a unit 2-sphere. At fixed $R$ we are dealing with a spherical shell of dust 
  particles with a radius evolving as $r(R,\tau)$. The SET of dustlike matter with density 
  $\rho$, in its comoving frame, has the only nonzero component $T^0_0 = \rho$, while 
  the SET of a radial \emag\ field has the form
\beq   									\label{SET-e}
		{T\mN}^{[{\rm em}]} = \frac{q^2}{8\pi G r^4} \diag(1,\ 1,\ -1,\ -1),
\eeq  
  in properly chosen units, where $q = \const$ is a charge to be interpreted, for certainty, 
  as a magnetic one.

  The Einstein equations can be written as 
\bearr
	   2r\ddot{r} + \dot{r}{}^2 +1 - e^{-2\lambda}r'{}^2 = \frac{q^2}{r^2},  \label{G11}
\\ \lal
		\frac{1}{r^2}(1 + \dot{r}^2 + 2 r\dot{r}\dot{\lambda})
       	- \frac{\e^{-2\lambda}}{r^2} (2rr'' + r'{}^2 - 2 rr'\lambda' )
            									= 8\pi G\rho + \frac{q^2}{r^4},  \label{G00}
\\ \lal
		      \dot{r}' - \dot{\lambda} r' = 0.								\label{G01}
\ear
  where dots denote $\d/\d\tau$ and primes $\d/\d R$. Integrating \eqn{G01}) in $\tau$,
  we obtain 
\beq                  \label{lam}
		\e^{2\lambda} = \frac{r'{}^2}{1 + f(R)}, 
\eeq
  with an arbitrary function $f(R) > -1$. With \rf{lam}, \eqn{G11} leads to
\beq 							\label{ddr}
  			  2r \ddot{r} + \dot{r}^2 = f(R) + \frac{q^2}{r^2},
\eeq
  and its first integral is
\beq 									 \label{dr}
			\dot{r}^2 = f(R) + \frac{F(R)}{r}- \frac{q^2}{r^2},
\eeq
  $F(R)$ being one more arbitrary function. From \rf{dr} it is clear that 
  the function $f(R)$ characterizes the initial velocity distribution of dust particles.
  With $f \geq 0$, a particle can reach arbitrary large radii $r$, so that $f >0$ 
  corresponds to hyperbolic motion and $f=0$ to parabolic motion.
  Accordingly, $f(R) < 0$ describes elliptic motion, at which a particle cannot 
  reach infinity, and a maximum accessible value of $r$ for given $R$ can be 
  found from \eqn{dr} by imposing $\dot r =0$.   

  The constraint equation \rf{G00}, after substitution of  \rf{lam} and \rf{dr},
  gives for the density 
\beq
		\rho  = \frac{1}{8\pi G} \frac{F'(R)}{r^2 r'},       \label{rho}
\eeq
  whence it follows
\beq  												\label{F-tol}
		F(R) = 2 G M(R) = 8\pi G \int \rho r^2 r' dR.
\eeq
  where $M(R)$ is called the mass function. In space-times with a regular
  center, integrating in \rf{F-tol} from this center to a given $R$, we obtain $M(R)$ 
  as the mass of a body inside the sphere with this $R$, and $F(R)$ is its \Scw\ radius.
  However, the relations \rf{rho} and \rf{F-tol} are still valid if a regular center is absent.
  Note that $F$, and $q$ have the dimension of length while $f$ is dimensionless. 
  
  The results of further integration of (\ref{dr}) in $\tau$ depend on the sign of $f(R)$:
\bearr                                            \label{tau+}
     f >0: \quad 
         \pm[\tau-\tau_0(R)] = \frac {\sqrt{fr^2 +Fr -q^2}}{f} 
         			- \frac F{f^{3/2}} \ln\!\Big(F +\! 2fr +\! 2\sqrt{f} \sqrt{fr^2 +\! Fr -\! q^2} \Big), 
\yyy                                         \label{tau0}
     f = 0: \quad 
     		 \pm[\tau-\tau_0(R)] = \frac{2 \sqrt{Fr -q^2} (F r + 2 q^2)}{3 F^2},
\yyy                                      \label{tau-}
     f < 0: \quad 
     		 \pm[\tau-\tau_0(R)] =  \frac 1h \sqrt{-hr^2 +Fr - q^2} 
     		 			+ \frac {F}{2 h^{3/2}} \arcsin \frac {F-2hr}{F^2 - 4hq^2}, .
\ear
  where for $f < 0$ we denote $- f(R) = h(R) >0$, and the arbitrary function $\tau_0(R)$ 
  corresponds to the choice of a zero point of $\tau$ on each comoving sphere 
  with fixed $R$. The elliptic model (\ref{tau-}) exists under the requirement that 
  the quadratic equation $hr^2 - Fr +q^2 =0$ for the maximum accessible radius 
  $r$ has real roots, whence $F^2 - 4hq^2\geq 0$

  If we put $F = 2GM = \const$ in this  solution, we obtain $\rho=0$ and the \RN\ metric 
  with mass $M$ and charge $q$, written in a geodesic reference frame, perfectly
  suitable for smooth matching with the internal solution with variable $F(R)$.
  It is therefore easy to obtain a global solution for a dust cloud surrounded by an 
  empty \RN\ region. A global configuration can be formed by the q-LTB solution with 
  variable $F(R)$ in some range of $R$ (for example, $R < R_0$) and constant outside it.
  
  This leads to an interpretation of the function $M(R)$ in any q-LTB solution,
  even without a center or with a singular center: $M(R)$ is the \Scw\ mass in 
  the external \RN\ or \Scw\ solution that can be joined and matched to the present 
  solution at this particular value of $R$.

\subsection{Possible throats in LTB space-times}

  Let us try to obtain a \wh\ model on the basis of the q-LTB solution. According to
  the above-said, we will define a throat as a minimum of $r(R,\tau)$ on spatial
  sections  $\tau = \const$ of our space-time. It means that on such a throat we 
  must have $r' =0$ and $r'' > 0$ in terms of some admissible $R$ coordinate, quite 
  similarly to the conditions used in \ssph\ metrics. (Note that the coordinate $r$ or 
  its multiple with a constant factor, used in many studies, is not admissible on a throat.) 
  
  In particular, manifestly admissible is the Gaussian coordinate $l$, equal to length 
  in the radial direction, such that $\e^{2\lambda} =1$ (which can always be 
  achieved at least locally). With $R=l$, according to \rf{lam}, $r'^2 = 1- f(R)$, and
  we have on the throat $R=R_0$
\beq                             \label{th1}
		  	r'^2 = 1+ f(R_0) = 0 \ \ \then \ \ f(R_0) = -1, \ \ \ {\rm or}\ \ \  h(R_0)=1.
\eeq
  It immediately follows that only elliptic models \rf{tau-} are compatible with throats.
  Next, to keep the metric \rf{ds-tol} nondegenerate, it must be $1+f  = 1 -h >0$, 
  and we conclude that $h = 1$ can only take place at a particular value of $R$, so
  that $R=R_0$ is a maximum of $h(R)$, whence $h'(R_0) =0$ and $h''(R_0) <0$. 
  
  Calculating $r''$ from \rf{lam} for $R=l$ (i.e., $\e^\lambda =1$), we obtain 
\beq                             \label{th2}
			r''\Big|_{R=R_0} = - \frac{h'}{2r'}\Big|_{R=R_0} > 0.
\eeq   
  Thus $h'$ should vanish at $R=R_0$ along with $r'$, and there ratio should be finite
  and negative.
  
   Requiring that the density should be positive, we should have $F'/r' > 0$ on the 
  throat, therefore, as $R$ changes, both $F'$ and $r'$ 
  should change their sign simultaneously. (Note that \whs\ in q-LTB space-times
  with negative matter density have been studied in \cite{sha08}.)
   
  Thus we can summarize that on a regular throat $R=R_0$ with $\rho >0$ it holds
\bearr                   		 \label{throat}
		r' =0,\quad\   h = 1, \quad\ h'=0, \quad\ h'' < 0; \quad\ \frac{h'}{r'} < 0;
\nnn		
		F' = 0, \quad\ \frac{F'}{r'} > 0, \quad\  F^2 - 4hq^2 > 0.
\ear
  
  Further on it will be helpful to use the well-known parametric presentation of
  the elliptic solution \rf{tau-} \cite{LL}: assuming the synchronization function 
  $\tau_0(R) \equiv 0$ in \rf{tau-}, we have
\bear                 			 		\label{eta}
			r \eql \frac {F}{2h} (1 - \Delta \cos \eta), 
\nn
			\tau \eql \frac {F}{2h^{3/2}}(\eta - \Delta \sin\eta),
						\qquad  \Delta = \sqrt{1 - \frac{4 hq^2}{F^2} },
\ear  
  where $0 < \Delta \leq 1$, and $\Delta =1$ corresponds to $q=0$, the LTB solution 
  with pure dust. This LTB solution possesses singularities $r=0$ at $\eta =0, 2\pi,$ etc., 
  while at $q \ne 0$ we have $\Delta < 1$, and the value $r=0$ is never achieved. 
  With $q=0$ ($\Delta =1$), under the assumptions 
\beq       \label{Fri1}
	    F(\chi) = 2 a_0 \sin^3\chi, \qquad   h (\chi) = \sin^2 \chi, \qquad a_0 = \const
\eeq  
  (using the radial coordinate $R = \chi$ of angular nature), the solution describes
  Friedmann's closed isotropic universe filled with dust \cite{LL}, such that
\beq         \label{Fri2}  
  		r = r(\eta,\chi) = a(\eta) \sin \chi,  \qquad a(\eta) = a_0 (1-\cos \eta),
\eeq  
  where $a(\eta)$ is the cosmological scale factor. 
  
  Let us return to possible throats in q-LTB space-times. By \rf{eta}, the derivative 
  $r'$ on a constant-$t$ spatial section of our space-time is given by
\bearr 
		r'=\frac{Fh'N_1(R,\eta) + 2h F' N_2(R,\eta)} {4\Delta h^2(1-\Delta\cos\eta)}, 
\nnn
		N_1(R,\eta) = \cos\eta -3\Delta + 3\Delta^2 (\cos\eta + \eta\sin\eta) 
							 + \Delta^3 ( -2 + \cos^2\eta),
\nnnv					
		N_2(R,\eta) = 	- \cos\eta + 2\Delta- \Delta^2(\cos\eta + \eta\sin\eta)
\ear   

  Consider the finite limit $\lim\limits_{R\to R_0} [F h'/(F'h)] = - B$, so that 
  $B = \const >0$. Then we have on such a throat 
\bearr                          \label{r_R}
		r'\Big|_{R_0} = \frac {2 N_2 - BN_1}{2 \Delta\, ( 1-\Delta\,\cos \eta)},
\yyy                           \label{rho_R0}
	\rho\big|_{R_0} = \frac{F'}{8\pi G r^2 r'}\bigg|_{R_0} 
	= \frac{\Delta ( 1-\Delta\,\cos \eta )}{2\pi G r^2 (2N_2 - BN_1)}\bigg|_{R_0}.
\ear
      
  One more important observation follows from a comparison of the q-LTB solution 
  with the definition of R- and T-regions. Indeed (see, e.g., \cite{BR21}), a T-region 
  is characterized by a timelike gradient of $r(R, \tau)$, an R-region by its spacelike
  nature, and this gradient is null at an apparent horizon. For the solution under study,
  by \rf{lam} and \rf{dr},
\beq                                                       \label{grad}
	   r^{,\alpha}r_{,\alpha} = \dot{r}^2 - \e^{-2\lambda}r'{}^2
			= -1 + \frac {F(R)}{r} - \frac {q^2}{r^2},
\eeq
  Since $F(R) = 2M(R)$, we see that the R- and T-regions are here quite similar to
  the \RN\ space-time, but the mass of a gravitating center is here replaced by 
  the current value of the mass function $M(R)$.  
   
  Substituting the condition \rf{th1} and $r(R, \tau)$ from \rf{eta}, we find that on a 
  throat in q-LTB space-time, we have
\beq                                                       \label{grad-th}
	r^{,\alpha}r_{,\alpha} = \frac 1{r^2} \sin^2 \eta \Big(\frac{F^2}{4} - q^2 \Big) \geq 0.
\eeq  
  Thus such a throat is, in general, located in a T-region, except for the instants
  $\eta = \pi n,\ n\in \N$ (at which $r(R,\tau)$ takes extremal values as a function of 
  $\tau$), at which the throat is located at a horizon. This means, in particular, 
  that matching of a q-LTB \wh\ configuration to a \RN\ external region cannot lead
  to an \asflat\ \wh\ but only to a \bh. Such models are studied in 
  \cite{sushkov20, we21}. The same q-LTB solution with an electric field, has 
  been used for a study of a possible structure of charged particles in \cite{sukhan18}.
  
\subsection{Wormholes in a dust-filled universe}

  Let us try to construct an example of a q-LTB \wh\ solution, choosing the following 
  simple functions of $R =: x$ in agreement with the requirements \rf{throat}:\footnote 
  		{We are using the letter $R$ for a general radial coordinate and $x$ for 
  		its specific choice.}
\beq                 \label{ex1}
		h(x)=\frac{1}{1+x^2}, \quad \ F(x)=2b (1+x^2), \ \then \  
		\Delta=\sqrt{1-\frac{q^2}{b^2(1+x^2)^3}}, 
\eeq
   ($b =\const > 0$), so that 
\beq                \label{ex1-r}
		r(x,\eta) = b (1+x^2)^2 (1 - \Delta \cos \eta),
				\qquad
		r'(x,\eta) = \frac{b x (1+x^2) (2N_2 - N_1)}{\Delta(1-\Delta \cos \eta)}, 
\eeq
  with $N_{1,2}$ defined in \eqn{r_R}. The throat is located at $x=0$, and the whole
  solution is symmetric with respect to it. Different signs of the derivatives of $h(x)$ 
  and $F(x)$, under the condition $N_2(x,\eta) >0$, provide $\rho > 0$ on the throat 
  and some its neighborhood, but the same is not guaranteed at all $x$ and $\eta$.
  The time evolution of the throat radius $r(0,\eta)$ is shown in Fig.\,2, 
  and the behavior of $r'$ outside the throat in Fig.\,3 (on the throat itself 
  we obviously have $r' \equiv 0$).  The quantity $b$ is an arbitrary length scale,
  and in calculations for all figures we have assumed $b=1$.
\begin{figure}[ht]
\centering
\includegraphics[scale=0.4]{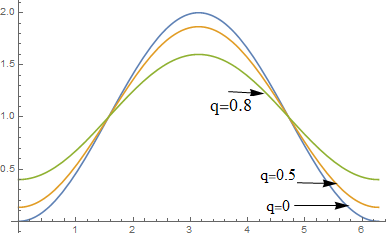}
\caption{\small
	Time dependence of the throat radius $r(0,\eta)$ for $q=0, 0.5, 0.8$}
\end{figure}
\begin{figure}[ht]
\centering
\includegraphics[scale=0.26]{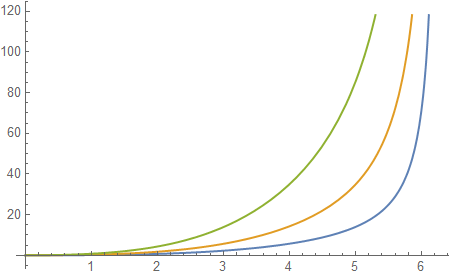} \ 
\includegraphics[scale=0.26]{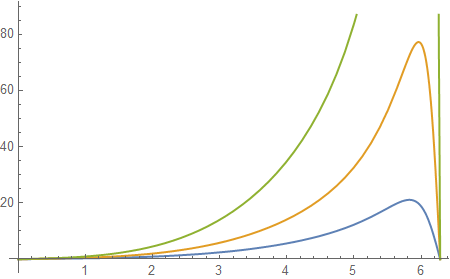}\
\includegraphics[scale=0.26]{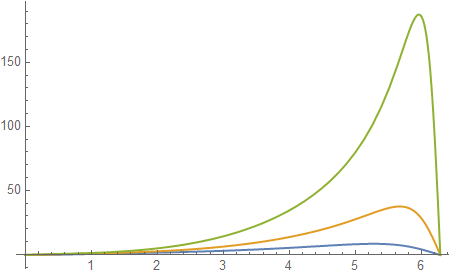}\
\caption{\small
		Time dependence of $r'(x,\eta)$ for $q=0$ (left), $q=0.5$ (middle)
		and $q=0.9$ (right).}.
\end{figure}

  It is of interest that the dependence of the proper time $\tau$ on the time parameter 
  $\eta$ is almost insensitive to a nonzero ``charge'' $q$, see Fig.\,4.
\begin{figure}[ht]
\centering
\includegraphics[scale=0.35]{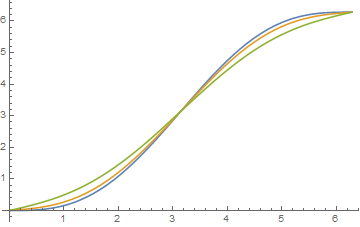}\ \
\includegraphics[scale=0.35]{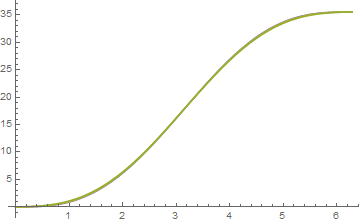}
\caption{\small
	Proper time $\tau$ in terms of $\eta$ for $q=0, 0.5, 0.8$ on the throat 
	$x=0$ (left, the plot is less steep for larger $q$) and $x=1$ (right, the plots 
	almost merge).}
\end{figure}
  
   As we already know, the equality $r'=0$, if not on a throat, means that there is a singularity
   with infinite matter density, and to have finite $\rho > 0$ we need $r' > 0$ at $x > 0$
   and $r' < 0$ at $x < 0$. As follows from calculations and is illustrated in Fig.\,3, 
   we observe a good \wh\ behavior of our solution at all $\eta \in (0, 2\pi)$.
   
   The density $\rho$ calculated according to \rf{rho} is also positive and well-behaved,
   as illustrated in Fig.\,5. The density diverges as $\eta^{-6}$ at $\eta \to 0$, i.e., at
   the beginning of the evolution, and as $(2\pi - \eta)^{-3}$ at its end ($\eta \to 2\pi$)
   for dust ($q=0$) but is everywhere and always finite for $q\ne 0$, with 
   qualitatively the same time dependence.
\begin{figure}[ht]
\centering
\includegraphics[scale=0.37]{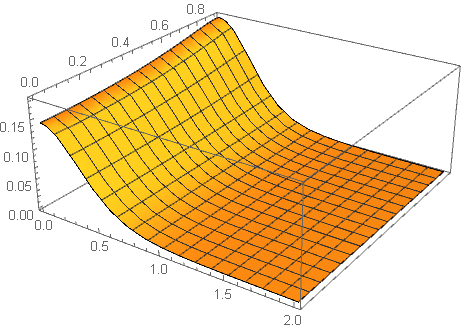}\ 
\includegraphics[scale=0.42]{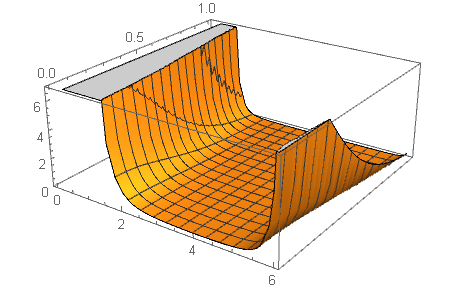}
\caption{\small
	The quantity $\rho/(8\pi)$ as a function of $x$ and $q$ for $\eta =\pi$ (left) 
	and as a function of $x$ and $\eta$ for $q=0$ (right). }
\end{figure}   

   Now, let us see how to inscribe such a \wh\ into the cosmological model \rf{Fri1},
   \rf{Fri2}. To match two q-LTB solutions with different choices of the arbitrary 
   functions $F(R)$ and $h(R)$ at some value $R = R^*$, one should first of all 
   identify the junction hypersurface as seen ``from the left'' and ``from the right,''
   to be treated as the same surface in unified space-time, hence, in the metric
   \rf{ds-tol} the coefficients $g_{\tau\tau} \equiv 1$ and $g_{\theta\theta} = r^2$
   must coincide. For $g_{\tau\tau}$ it is trivial, while the requirement to the jump 
   $[r^2] =0$ is meaningful. Next, to avoid the emergence of a thin shell of matter 
   on the junction surface, its second quadratic form should also have no jump
   \cite{darmois, israel}. For the metric \rf{ds-tol}, this requirement reduces to 
   $[\e^{-\lambda} r'] =0$, while a similar requirement to $g_{\tau\tau}$ is again 
   trivial. With \eqs \rf{lam} and \rf{eta}, the above requirements take the form
\beq                          \label{junc}
           [r] = 0, \quad\  [\e^{-\lambda} r'] =0 \ \ \then \ \ [h] =0, \quad [F]=0.
\eeq    
  In other words, to match two particular solutions, we must simply identify 
  the values of $F(R)$ and $h(R)$ on the junction surface. The charge $q$ 
  should certainly be the same in both solutions to provide continuity of the 
  \emag\ field.  It is of utmost importance that due to \rf{eta} these matching 
  conditions hold at all times as long as both solutions remain regular. 
  There is no need to adjust the radial coordinate choice in the two solutions since 
  both $r$ and $\e^{-\lambda} r'$ are insensitive to this choice, and consequently 
  the same is true for $h(R)$ and $F(R)$ which behave as scalars under 
  transformations of $R$. 
  
  Let us apply the conditions \rf{junc} to the solutions \rf{Fri1}, \rf{Fri2}
  and \rf{ex1}, \rf{ex1-r}, in which we must put $q=0$. If the junction surface
  is specified by $\chi = \chi^*$ and $x = x^*$, we obtain
\beq                    \label{junc1}
			x^* = \cot \chi^*, \qquad b = a_0 \sin^5 \chi^*.
\eeq   
  This ensures matching at $x^* > 0$. With the even functions \rf{ex1}, the 
  same matching is implemented at $x^* < 0$, and the whole configuration 
  consists of two closed evolving Friedmann universes connected through a \wh,
  and one can imagine something like a dumbbell (replacing 3D spherical spaces 	   
  with 2D spherical surfaces). 
  
  Some numerical estimates are in order. Assuming $a_0 \sim 10^{28}$ cm, 
  approximately the size of the visible part of our Universe, let us require that the 
  throat radius $r_{\rm th} \sim b$ should be much larger that the Planck length, 
  $b \gg l_{\rm pl} \sim 10^{-33}$ cm. Then from \rf{junc1} it is easy to obtain 
\beq   
			\sin \chi^* \gg 10^{-12}, \qquad   x^* \ll 10^{12}, 	
\eeq
  so that $r^* = r(x^*) \gg 10^{16}$ cm $\sim 0.01$ light year. Thus the wormhole 
  region must be large enough on the astronomical scale in order to provide a reasonable 
  size of the throat. Assuming $r^* \sim 100$ light years, the size of a small stellar cluster,
  we obtain $\chi^* \sim 10^{-8}$ and $r_{\rm th} \sim 10^{20} l_{\rm pl}$, approximately 
  the nuclear size of the throat.  Assuming a galactic size of the \wh\ region, 
  $r^* \sim 30\ {\rm kpc} \sim 10^{23}$\,cm, we have $\chi \sim 10^{-5}$ 
  and consequently $r^* \sim 10^3$ cm, a macroscopic size of the throat.
  
  The density obeys \eqn{rho}, which for our model \rf{ex1} and $q=0$ gives 
\beq                       \label{rho1}
  		\rho = \frac{1}{2\pi G b^2}\, \frac{1}{(1+x^2)^5 (1-\cos\eta)(2N_2 - N_1)}
\eeq   
  where $2N_2 - N_1$ is, for $\Delta =1$, a function of $\eta$, positive in the 
  range $\eta \in (0, 2\pi)$. The time-dependent part of \rf{rho1} may be approximated 
  by $1/10$, which is true for a larger part of the time range, and then we obtain, 
  by order of magnitude, 
\beq
             \rho \sim \frac {10^{-30}} {(1+x^2)^5 \sin^{10}\chi^*} \ \rm g/cm^3. 
\eeq  
   For the throat $x=0$ this expression gives about $10^{50}\ \rm g/cm^3$ if
   $\chi^* = 10^{-8}$ and about $10^{20}\ \rm g/cm^3$ if $\chi^* = 10^{-5}$, a
   value still substantially exceeding the nuclear density. On the other hand, since 
   $x^* \sim 1/\chi^*$, for the density at the junction surface $x =x^*$ we obtain a
   universal value of $\sim 10^{-30}  \ \rm g/cm^3$, close to the mean cosmological 
   density, independently of the $\chi^*$ value.
   
   Such estimates will certainly change, though probably not too drastically, if 
   we choose $F(R)$ and $h(R)$ other than in \rf{ex1}.
   
   The same procedure applied to solutions with $q\ne 0$ would describe a \wh\
   joining two Friedmann universes deformed by the magnetic field, but their 
   discussion is beyond the scope of this paper.     

\section{Wormholes leading to extra dimensions}

   The studies of multidimensional \whs\ are as diverse as are multidimensional extensions of 
   GR. Therefore, let us here, without trying to review this vast area of research, 
   which can be a subject of a much larger project, simply refer to a number of relevant papers 
   \cite{cle84, bobs, dzhu00, bene03, wh-bw, lobo07, dzhu14, dotti14, kuh18, baruah19},
   where further references can be found. In this section we consider a case of interest 
   where the size of compact extra dimensions is small enough at one entrance to a \wh\ 
   to make them invisible and is much larger at the other entrance \cite{ex-D1, ex-D2}.
   Such solutions to the field equations thus describe a transition from an effectively 4D 
   geometry to a multidimensional space-time.

\subsection{Field equations} 

  Consider 6-dimensional GR with a minimally coupled scalar field $\phi$ with a potential 
  $V(\phi)$ as the only source of gravity. The Lagrangian is 
\beq         \label{Lag}
             L = R_6 + 2\eps_\phi g^{AB} \d_A\phi \d_B\phi - 2V(\phi),
\eeq
  where $R_6$ is the 6D Ricci scalar, $\eps_\phi$ is $+1$ for a canonical scalar field and 
  $-1$ for a phantom one, and $A, B, \ldots = \overline{0,5}$. The equations of motion 
  include the scalar field equation $2\eps_\phi \Box_6\phi + dV/d\phi =0$ 
  ($\Box_6 = \nabla^A \nabla_A$ is the 6D d'Alembertian) and the Einstein equations 
  which may be written in the form
\bearr             \label{EE6}
             R^A_B = - \tT^A_B \equiv - T^A_B - \tfrac 14 \delta^A_B T^C_C 
                         = - 2\eps_\phi  \d^A\phi \d_B\phi + \half V(\phi) \delta^A_B,
\ear
  $R^A_B$ being the 6D Ricci tensor and $T^A_B$ the scalar field SET.

  Let the 6D space-time be a direct product of three 2D subspaces,
  $\M = \M_0 \times \M_1 \times \M_2$, where $\M_0$ is 2D Lorentzian space-time 
  parametrized by the coordinates $x^0 = t$ and $x^1 =x$, while $\M_1$ and 
  $\M_2$ are compact spaces of nonnegative constant curvature, i.e., each of them 
  is either a sphere or a torus. The 6D metric is assumed in the form:
\beq                                                 \label{ds_6}
	ds^2 = A(x) dt^2 - \frac{dx^2}{A(x)} - R(x) d\Omega_1^2 - P(x) d\Omega_2^2,
\eeq 
  where $A(x),\ R= r^2(x),\ P= p^2(x)$ depend on the ``radial'' coordinate $x$, 
  satisfying the condition $g_{tt}g_{xx}= -1$ (the ``quasiglobal gauge'' \cite{BR21}),
  while $d\Omega_1^2$ and  $d\Omega_2^2$ are $x$-independent metrics 
  on the manifolds $\M_1$ and $\M_2$ of unit size. It is also assumed $\phi = \phi(x)$.    

  We do not specify which of the subspaces $\M_{1,2}$ belongs to our ``external'' 4D 
  space-time and which is ``extra.'' Thus, if $\M_1$ is large and spherical while $\M_2$ 
  is small and toroidal, we have a \ssph\ 4D space-time with a toroidal extra space;
  if $M_1$ is small and spherical and $\M_2$ is large and toroidal, then we are dealing 
  with toroidally symmetric 4D space-time and a spherical extra space, and so on.
  
  The scalar field equation can be derived from \rf{EE6}, and among the latter there are 
  four independent ones, which may be written in the form
\bear                                  		\label{00}
             R^t_t = - \tT^t_t \ &\then & \ - (PR)^{-1}(A' PR)' = V(\phi),
\\  	                                 		\label{01}     
             R^t_t - R^x_x = -\tT^t_t +\tT^x_x  \ &\then & \
				\frac{r''}{r} + \frac{p''}{p} = -\eps_\phi  \phi'{}^2,
\\  	                                 		\label{02}     
             R^t_t - R^{\underline a}_a = 0   \ &\then &\ [P (AR' - A'R)]' = 2\eps_1 P, 
\yy                                    		\label{05}
             R^t_t - R^{\underline m}_m = 0  \ &\then & \ [R (AP' - A'P)]' = 2\eps_2 R,
\ear
  where the prime denotes $d/dx$, the index $a= 2,3$ (belonging to $\M_1$), 
  while $m=4,5$ (belonging to $\M_2$), and there is no summing over an underlined index; 
  furthermore, $\eps_1 = 1$ if $\M_1$ is a sphere and $\eps_1 =0$ if it is a torus, 
  and similarly for $\eps_2$ and $\M_2$.
 
  Equations \rf{02} and \rf{05} can be considered separately as two equations for 
  the three metric functions $A(x), P(x),  R(x)$, so that there is arbitrariness in one function. 
  Further, if we know the metric functions, we can use the other two equations, \rf{00} 
  and \rf{01}, to find the scalar field $\phi$ and the potential $V$. 
  
  As follows from \eqn{01}, if we assume that $r > 0$ and $p > 0$ in the whole range 
  $x \in \R$, this is only compatible with $\eps_\phi = -1$, a phantom scalar, 
  otherwise at least one of these radii will turn to zero at some finite $x$. 

\subsection {Possible asymptotic behavior of the metric}

  The following types of geometry are possible with the metric \rf{ds_6}:
\begin{description}
\item[(i)]\ \ 
	 SS (double spherical) space-times if $\eps_1 = \eps_2 =1$.
\item[(ii)]\ 
	 ST (spherical-toroidal) space-times if $\eps_1 =1,\ \eps_2 =0$ or vice versa. 
\item[(iii)]
	 TT (double toroidal) space-times if $\eps_1 = \eps_2 =0$.
\end{description}
  For any of these types, we are interested in configurations where $x \in \R$ and 
  where one of the subspaces $\M_1$ or $\M_2$ is large on both ends $x\to \pm\infty$
  while the size of the other is radically different. In particular, a 4D flat asymptotic 
  region times small extra dimensions as $x\to -\infty$ and something different on the 
  other end. 
  
\def\fin{{\rm finite}}  
  
  Some restrictions can be obtained from the analysis of \eqs \rf{02} and \rf{05}, 
  without addressing the scalar field properties. Consider, for example, a 4D \asflat\ 
  space-time with everywhere finite extra dimensions in SS geometry. It means that
\beq                                                                                                  \label{flat+}
	A(x) \to \fin, \quad\ R(x) \sim x^2, \quad\ P(x) \to \fin \quad {\rm as}\ \ x\to \infty.
\eeq
  Let us use these conditions in \eqs \rf{02} and \rf{05}. From \rf{flat+} it follows that 
  $R' \sim x$, $A' \sim x^{-3}$ or even smaller (since for a finite limit of $A$ it  should be
  $A = A_- + A_{-2}/x^2 + \ldots$), and the l.h.s. of \rf{02} tends, in general, to a nonzero 
  constant, which conforms to the assumed limit of $P$ on the r.h.s.. However, 
  the expression in square brackets in \eqn{05} tends to a constant, its derivative thus
  vanishes, whereas the r.h.s. is $2R \sim x^2$. We have to conclude that 
  {\it the conditions \rf{flat+} contradict the field equations.}

  The same reasoning applies to $x\to -\infty$ and/or exchanged $R(x)$ and $P(x)$. 
  Other opportunities can be tested in the same manner. An analysis shows 
  \cite{ex-D2} that SS space-times cannot possess a flat Minkowski asymptotic 
  region times a sphere of finite size, while a de Sitter behaviour times a finite sphere 
  can take place. In ST geometry, we can obtain effectively 6D asymptotics (that is,
  $R$ and $P$ equally large), as well as an \asflat\ \sph\ 4D space-time times constant 
  extra dimensions. In general, one can obtain similar or different asymptotic behaviors
  at the two ends, $x \to \pm\infty$. We will present two examples of such solutions 
  to \eqs \rf{00}--\rf{05} with a 4D geometry with small extra dimensions at one end
  and an effectively 6D geometry at the other.

\subsection{Example 1: ST geometry, wormholes with a massless scalar}  

  An \asflat\ \sph\ geometry with small extra dimensions at one end and the same with much 
  larger extra dimensions at the other can be found among known solutions for a 
  massless scalar field ($V=0$) \cite{k95, bim97}. The solution of interest has the form
  \cite{ex-D1} 
\bearr                                       \label{ds_2}
       ds^2 = dt^2 - \e^{-4nu}\big[dz^2 + (z^2+k^2) d\Omega_1^2\big] -\e^{2nu}d\Omega_2^2,
\nnn
         \phi = Cu \equiv (C/k)\cot^{-1} (-z/k),	
\ear
  where $z \in \R$, $n > 0,\ C,\ k>0$ are integration constants connected by the relation
  $2C^2 = k^2 + 3 n^2$. In the 4D subspace $\M_0 \times \M_1$ it represents a \sph, 
  twice \asflat\ wormhole, while the extra subspace $\M_2$ is toroidal, with its 
  coordinates $x^4, x^5$ ranging from zero to some fixed length $a$. Thus  
  at $z = -\infty$ (hence $u=0$) the size of extra dimensions is equal to $a$, while at
  the other end, $u = \pi/k$, corresponding to $z = + \infty$, we have the size 
  $a_+ = \e^{n\pi/k} a$. In the trivial case $n=0$ the solution reduces to the Ellis 4D 
  \wh\ \cite{k73, hell} times an extra 2D toroidal space of constant size.

  Let us suppose that the size $a$ of extra dimensions at the left end, $z=-\infty$, 
  is sufficiently small to be invisible for the existing experimental means, say, 
  $a = 10^{-17}$ cm. The size $a_+$ corresponding to $z\to \infty$ depends on the 
  ratio $n/k$ and can be arbitrarily large. Thus, we obtain $p_+ \sim 1$ m
  if we take $n/k \approx 14$. To get a size of stellar order, 
  $p_+ \sim 10^6\ {\rm km} = 10^{11}$ cm, we should assume $n/k \approx 20.5$.

  The \wh\ throat is found at a minimum of the 4D radius $r$ 
  ($g_{22} = \e^{-4nu}(z^2+ k^2)$), which occurs at $z = 2n$ and is given by
\beq               \label{r_thr} 
		r_{\min} = \sqrt{k^2 + 4n^2} \exp{\Big(\frac{2n}{k} \cot^{-1} \frac{2n}{k}\Big)}.
\eeq   
  This radius can be made sufficiently large by properly choosing the parameters
  $k$ and $n$. Thus, if $n/k \gg 1$ (as required for obtaining large $a_+$),
  we have $r_{\min} \approx 2 n \e \approx 5.4 n$, and the lengths $n$ and $k$ 
  should be very large as compared to $a$. For example, to have the throat size
  $r_{\min} = 10$ m, large enough to transport macroscopic bodies, we should
  take $n \sim 2\ {\rm m} = 2\times 10^{19} a$.
  
\subsection{Example 2: ST geometry, asymptotically AdS wormholes}

  Solutions to Einstein-scalar equations with nonzero potentials $V(\phi)$ can be found, 
  in most cases, only numerically. For our system \rf{00}--\rf{05}, there is an exception: 
  assuming $\eps_2=0$, integrating \eqn{05} and putting the emerging integration
  constant equal to zero, we obtain $P = cA$, $c = \const $, and \eqn{02} takes the form
  $A^3 (R/A)']' = 2A$, that is, one equation for two functions $A(x)$ and $R(x)$. 
  It may be rewritten as
\beq                      \label{R'}
	      \Big(\frac RA\Big)' = \frac {2}{A^3} \int A(x) dx.
\eeq   
  and is thus solved by quadratures if the function $A(x)$ is specified.
  Let us try to obtain a configuration where the 4D subspace is \asflat\ on the left end 
  and tends to AdS behavior on the right. We thus suppose $A \to 1$ as $x\to -\infty$ 
  and $A \sim x^2$ as $x \to +\infty$. It is rather hard to invent $A(x)$ with such 
  asymptotic properties that would yield more or less simple analytic expressions 
  for other quantities. Therefore, we have obtained a desirable example 
  by choosing a piecewise smooth function $A(x)$ \cite{ex-D1}:
\beq             \label{A(x)}    
               A(x) = 1,\ \  x \leq 0;\qquad A(x) = 1 + 3 x^2/a^2, \ \ x\geq 0 \qquad  (a = \const > 0). 
\eeq
  The resulting metric is $\rm C^1$ smooth, but there are jumps in the functions 
  $\phi'(x)$ and $V(x)$ (Fig.\,6). As mentioned above, the relation $P = cA$ holds 
  at all $x$, while $R(x)$ is given by 
 \beq                                               \label {R-}
             R(x) = x^2 + b^2 \ \  (x \leq 0),\qquad
             R(x) = \Big(1 + \frac{3 x^2}{a^2}\Big) \biggl[b^2 
                                    + \frac{x^2 (1 + 2 x^2/a^2)}{(1 + 3 x^2/a^2)^2}\biggr] \ \  (x \geq 0),
\eeq  
  with $b = \const > 0$ (so that $x=0$ is a throat of radius $b$). At $x < 0$ the
  scalar field is massless, $V(x) \equiv 0$, and $\phi(x) = \arctan(x/b)$, 
  while at $x > 0$ the corresponding expressions are rather bulky.
\begin{figure}
\centering
\includegraphics[width=5.5cm]{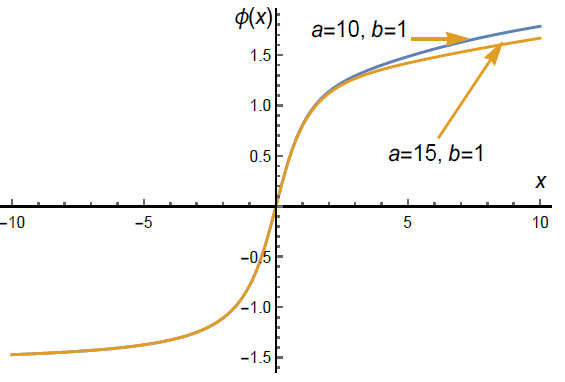}
\qquad
\includegraphics[width=5.5cm]{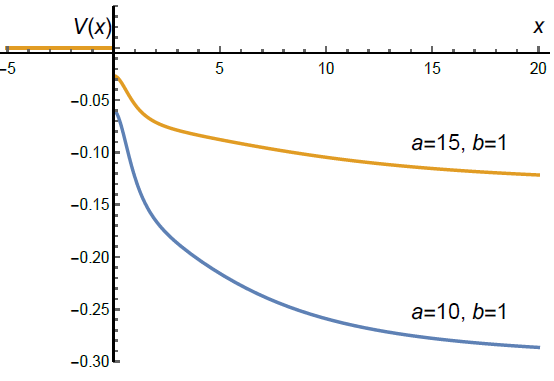}
\caption{\small 
	The scalar field $\phi(x)$ (left) and the potential $V(x)$ (right) in example 2}
\end{figure}
  
  We obtain a configuration which is \asflat\ times constant (arbitrarily small) 
  extra dimensions on one side of the \wh\ and behaves as a 6D AdS space on the 
  other side, with the potential $V$ tending to a negative constant value, becoming there
  an effective cosmological constant. 
  
  Let us note that it is easy to remove the jumps in $V(x)$ and $\phi'(x)$ at $x=0$ 
  by adding arbitrarily small quantities to $A(x)$ making it smoother than $\rm C^1$.
 
\section{Conclusion}

  We have considered three different kinds of \whs\ in GR, showing some possible 
  nonstandard features of these hypothetic objects. 
  
  The existence of stationary \cy\ \whs\ respecting the WEC not only shows a way to
  circumvent topological censorship but also illustrates the exotic properties of vortex
  gravitational fields due to rotation, see \eqn{vortex}. Similar solutions in the 
  cosmological context were obtained and discussed in \cite{azreg20}. The results of 
  \cite{radu21, kono21}, where \sph\ \whs\ were obtained with spinor sources of 
  gravity, probably show that the spin of matter can also manifest exotic properties,
  and this issue deserves a further study. One can notice that topology alone, without 
  rotation, cannot lead to viable phantom-free \wh\ solutions, as follows, in particular, 
  from the properties of static \cy\ \whs\ \cite{cyl-wh} and the recently obtained 
  static toroidal \whs\ \cite{dz-tor}.
  
  Cosmological \whs\ can probably be relevant in the early Universe, and the presently 
  reported results show that they could naturally exist not only at the inflationary stage
  but also at the matter-dominated one. It supports the idea that \whs\ with de Sitter 
  asymptotic regions that existed at the inflationary stage might somehow survive at the 
  radiation-dominated and later stages and still exist in the present Universe, as is 
  discussed by a number of researchers, e.g., in 
  \cite{roman93, kard06, sushkov07, fara10, savel15} and others.
  
  In 6D GR with a minimally coupled phantom scalar as a source of gravity, we have 
  constructed examples of space-times which are effectively 4D in a certain part of the
  whole space-time and effectively 6D on the far end. The existence of such ``bubbles''
  or their analogues with a different number of extra dimensions, surrounded by domain 
  walls, in our Universe cannot be a priori excluded, as shown, in particular, in \cite{extra21}, 
  and their possible observational consequences can be a subject of further studies. 
  One of such consequences can be \bh\ formation due to domain wall shrinking,
  which, if these ``bubbles'' were numerous enough in the early Universe, could result in a 
  wealth of primordial \bhs\ substantially contributing to the Dark Matter density \cite{extra21}. 
  As follows from our discussion, the anomalous domains can communicate with those  
  of conventional 4D physics not only by photons or gravitational waves crossing
  the domain walls, but also through \whs.  



\paragraph{Acknowledgments.}
	{I thank Milena Skvortsova, Sergei Bolokhov, Sergey Sushkov and Pavel Kashargin
	for collaboration and helpful discussions.
	This publication was supported by the RUDN University Strategic Academic 
	Leadership Program. It was also funded by the Ministry of Science and Higher 
	Education of the Russian Federation, Project ``Fundamental properties of elementary 
	particles and cosmology'' N 0723-2020-0041, and by RFBR Project 19-02-00346.
	}





\small
\begin{thebibliography}{99}\itemsep 1pt

\bibitem{flamm}
	L. Flamm, 
	Beitr\"age zur Einsteinschen Gravitationstheorie,
	Phys. Z. {\bf 17}, 448 (1916).

\bibitem{ER}
	A. Einstein and N. Rosen, The particle problem in the general theory of relativity, 
	Phys. Rev. {\bf 48}, 73-77 (1935).

\bibitem{rev1}
	M. Visser, {\it Lorentzian Wormholes: From Einstein to	Hawking }
	(Springer-Verlag, Berlin, 1997).

\bibitem{rev2}
	F.S.N.  Lobo, Exotic solutions in general relativity: Traversable wormholes and 
	``warp drive'' spacetimes. In: {\it Classical and Quantum Gravity Research}
	(Nova Science Publishers, NY, 2008), p. 1--78.	
	
\bibitem{BR21}
	K.A. Bronnikov and S.G. Rubin, 
	{\it Black Holes, Cosmology and Extra Dimensions}
	(2nd edition, World Scientific, Singapore, 2021).

\bibitem{censor1}
	J.L. Friedman, K. Schleich and D.M. Witt, Topological Censorship, 
	Phys. Rev. Lett. {\bf 71}, 1486 (1993); 
	Erratum: Phys. Rev. Lett. {\bf 75}, 1872(E) (1995).

\bibitem{censor2}
	G.J. Galloway, On the topology of the domain of outer	communication, 
	Class. Quantum Grav. {\bf 12}, L99 (1995).
     
\bibitem{cyl1}
	K.A. Bronnikov and V.G. Krechet,
	Potentially observable cylindrical wormholes without exotic matter in GR,
	Phys. Rev. D {\bf 99}, 084051 (2019); arXiv: 1807.03641.

\bibitem{cyl2}
	S.V. Bolokhov, K.A. Bronnikov, and M.V. Skvortsova,
	Cylindrical wormholes: A search for viable phantom-free models in GR,
	Int. J. Mod. Phys. D {28}, 1941008 (2019); arXiv: 1903.09862.

\bibitem{cyl3}
	K.A. Bronnikov, V.G. Krechet and V.B. Oshurko,
	Rotating Melvin-like universes and wormholes in general relativity,
	Symmetry {\bf 2020}, 12, 1306 (2020); arXiv: 2007.01145.	

\bibitem{wh-dS}
	K.A. Bronnikov, K.A. Baleevskikh and M.V. Skvortsova,
	Wormholes with fluid sources: A no-go theorem and new examples,
	Phys. Rev. D {\bf 96}, 124039 (2017); arXiv: 1708.02324. 
	
\bibitem{sushkov20} 
	P. Kashargin and S. Sushkov, 
	Collapsing wormholes sustained by dustlike matter, 
	Universe {\bf 6} (10), 186 (2020).
	
\bibitem{we21} 
	K.A. Bronnikov, P. Kashargin and S.V. Sushkov, 	
	Magnetized dusty black holes and wormholes, 
	in preparation.	
	
\bibitem{lemaitre}
	G. Lema\^{\i}tre, L'Univers en expansion,
	Ann. Soc. Sci. Brussels A {\bf 53}, 51 (1933);
	reprinted: Gen. Rel. Grav. {\bf 29}, 641 (1997).

\bibitem{tolman}	
	R.C. Tolman, Effect of inhomogeneity on cosmological models, 
	Proc. Nat. Acad. Sci. USA {\bf 20}, 169 (1934);
	reprinted: Gen. Rel. Grav. {\bf 29}, 931 (1997).

\bibitem{bondi}
	H. Bondi, Spherically symmetrical models in general relativity,
	Mon. Not. R. Astron. Soc. {\bf 107}, 410 (1947).
	reprinted: Gen. Rel. Grav. {\bf 31}, 1783 (1999).	
	
\bibitem{ex-D1}
	K.A. Bronnikov and M.V. Skvortsova, 
	Wormholes leading to extra dimensions. Grav. Cosmol. {\bf 22}, 316 (2016).
	
\bibitem{ex-D2}	
	K.A. Bronnikov, P.A. Korolyov, A, Makhmudov and M.V. Skvortsova, 
	Wormholes and black universes communicated with extra dimensions,
	J. Phys. Conf. Series {\bf 798}, 012091 (2017).
		
\bibitem{thorne}
	M. Morris, K. S. Thorne, and U. Yurtsever, 
	Wormholes, time machines, and the Weak Energy Condition, 
	Phys. Rev. Lett. {\bf 61}, 1446 (1988).

\bibitem{hoh-vis}
	D. Hochberg and M. Visser, 
	Geometric structure of the generic static traversable wormhole throat, 
	Phys. Rev. D {\bf 56}, 4745 (1997).

\bibitem{zipoy}
	D. Zipoy, Topology of some spheroidal metrics. 
	J. Math. Phys. {\bf 7}, 1137-1143 (1966).
	
\bibitem{ax1}
	K.A. Bronnikov and J.C. Fabris,
	Weyl spacetimes and wormholes in D-dimensional Einstein and 
	dilaton gravity, Class. Quantum Grav. {\bf 14}, 831-842 (1997).

\bibitem{ax2}
	G\'erard Cl\'ement,
	Spinning ring wormholes: a classical model for elementary particles?
	arXiv: gr-qc/9810075.

\bibitem{ax3}	
	A. Burinskii,
	Stringlike structures in the real and complex Kerr-Schild geometry,
	J. Phys. Conf. Ser. {\bf 532}, 012004 (2014); arXiv: 1410.2462.
	
\bibitem{cyl-EM}	
	K.A. Bronnikov, Static, cylindrically symmetric Einstein-Maxwell fields, 
	in {\it Problems in gravitation theory and particle theory (PGTPT)} 
	(Ed. K.P. Staniukovich, 10th issue, p. 37--50, Atomizdat, Moscow, 1979, in Russian).

\bibitem{cyl-wh}	
	K.A. Bronnikov and Jos\'e P.S. Lemos,
	Cylindrical wormholes,
	Phys. Rev. D {\bf 79}, 104019 (2009).

\bibitem{cyl-rev}
	K.A. Bronnikov, Nilton Santos and Anzhong Wang,
	Cylindrical systems in general relativity (review).
	Class. Quantum Grav. {\bf 37}, 113002 (2020); arXiv: 1901.06561.

\bibitem{cyl-rot}
	K.A. Bronnikov, V.G. Krechet, and Jos\'e P.S. Lemos,
	Rotating cylindrical wormholes. 
	Phys. Rev. D {\bf 87}, 084060 (2013); arXiv: 1303.2993.

\bibitem{krechet}	
	V.G. Krechet and D.V. Sadovnikov, 
	Spin-spin interaction in general relativity and induced geometries with
	nontrivial topology, Grav. Cosmol. {\bf 15}, 337 (2009).
	
\bibitem{darmois}	
	G. Darmois, Les \'equations de la gravitation einsteinienne. 
	In: {\it M\'emorial des Sciences Mathematiques, vol. 25} 
	(Gauthier-Villars, Paris, 1927).
	
\bibitem{israel}		
	W. Israel, Singular hypersurfaces and thin shells in general relativity. 
	Nuovo Cim. B {\bf 48}, 463 (1967).

\bibitem{phi-nogo}
	K.A. Bronnikov, Rotating cylindrical wormholes: A no-go theorem. 
	J. Phys. Conf. Series  {\bf 675}, 012028 (2016); arXiv: 1509.06924.

\bibitem{exact-book}
	H. Stephani, D. Kramer, M.A.H. MacCallum, C. Hoenselaers, E. Herlt, 
	{\it Exact Solutions of Einstein's Field Equations}, 
	Cambridge Monographs on Mathematical Physics
	(Cambridge University Press, 2009).
	
\bibitem{safko72}	
	J. Safko and L. Witten, 
	Some properties of cylindrically symmetric gravitational field,
	Phys. Rev. D {\bf 5}, 293--300 (1972).
	
\bibitem{evans77}	
	A.B. Evans,  Static fluid cylinders in general relativity, 
	J. Phys. A: Math. Gen. {\bf 10}, 1303--1311 (1977).
	
\bibitem{cyl-flu1}  
	K.A. Bronnikov, Static fluid cylinders and plane layers in general relativity,
	J. Phys. A: Math. Gen. {\bf 12},  201--207 (1979).
	
\bibitem{cyl-flu2}  
	K.A. Bronnikov, V. Abdel-Sattar, E.N. Chudaeva and G.N. Shikin, 
	Static, cylindrically symmetric perfect fluid configurations,
	Vestnik RUDN {\bf 1}, 85--95	(2009).
	
\bibitem{stiff-w} 
	K.A. Bronnikov, Gravitational and sound waves in stiff matter,
	J. Phys. A: Math. Gen. {\bf 13}, 3455--3463 (1980). 	
	
\bibitem{santos82}	
	N.O. Santos and R.P. Mondaini, 
	Rigidly rotating relativistic generalized dust cylinder. 
	Nuovo Cim. B {\bf 72}, 13 (1982).
	
\bibitem{sklav99}	
	D. Sklavenites, Stationary perfect fluid cylinders, 
	Class. Quantum Gravity {\bf 16}, 2753 (1999).

\bibitem{ivanov02}	
	B.V. Ivanov, On rigidly rotating perfect fluid cylinders, 
	Class. Quantum Gravity {\bf 19}, 3851 (2002).
	
\bibitem{santos06}
     F. Debbasch, L. Herrera, P.R.C.T. Pereira, and N.O. Santos,
     Stationary cylindrical anisotropic fluid,
     Gen. Rel. Grav. {\bf 38}, 1825 (2006); gr-qc/0609068.	
		
\bibitem{aniso}                   
	S.V. Bolokhov, K.A. Bronnikov, and M.V. Skvortsova.
	Rotating cylinders with anisotropic fluids in general relativity.
	Grav. Cosmol. {25} 122--130 (2019); arXiv: 1904.06727.

\bibitem{stiff1}
	S.D. Odintsov, V.K. Oikonomou,
	The early-time cosmology with stiff era from modified gravity,
	arXiv: 1711.04571
	 	
\bibitem{stiff2}
	G. Brando, J.C. Fabris, F.T. Falciano, O. Galkina,
	Stiff matter solution in Brans-Dicke theory and the general relativity limit,
 	arXiv: 1810.07860 	
		
\bibitem{berezin87}
	V.A. Berezin, V.A. Kuzmin, and I.I. Tkachev, 
	Dynamics of bubbles in general relativity,
	Phys. Rev. D {\bf 36}, 2919 (1987). 

\bibitem{vilenk}
	A. Vilenkin, Gravitational Field of Vacuum Domain Walls and Strings, 
	Phys. Rev. D {\bf 23}, 852 (1981).
		
\bibitem{radu21}		
	Jose Luis Bl\'azquez-Salcedo, Christian Knoll, and Eugen Radu,
	Traversable wormholes in Einstein-Dirac-Maxwell theory,
	Phys. Rev. Lett. {\bf 126}, 101102 (2021); arXiv: 2010.07317.
		
\bibitem{kono21}
	R. A. Konoplya and A. Zhidenko,
	Traversable wormholes in general relativity without exotic matter,
	arXiv: 2106.05034.
	
\bibitem{comment}
	Kirill Bronnikov, Sergey Bolokhov, Serguey Krasnikov, and Milena Skvortsova,
	Comment on ``Traversable wormholes in Einstein-Dirac-Maxwell theory,''
	arXiv: 2104.10933.
	
\bibitem{bouh21}	
	Mariam Bouhmadi-L\'opez, Che-Yu Chen, Xiao Yan Chew, Yen Chin Ong and Dong-han Yeom,
	Traversable wormhole in Einstein 3-form theory with self-interacting potential.
	arXiv: 2108.07302.

\bibitem{hoch98}	
	D. Hochberg and M. Visser, 
	Dynamic wormholes, antitrapped surfaces, and energy conditions,
	Phys. Rev. D {\bf 58}, 044021 (1998).

\bibitem{hay99}
	S. A. Hayward, Dynamic wormholes, 
	Int. J. Mod. Phys. D {\bf 8}, 373 (1999); gr-qc/9805019.
	
\bibitem{tomikawa15}
	Y. Tomikawa, K. Izumi and T. Shiromizu, 
	New definition of a wormhole throat, 
	Phys. Rev. D {\bf 91}, 104008 (2015).	
	
\bibitem{bittencourt17}	
	E. Bittencourt, R. Klippert, and G.B. Santos,			
	Dynamical wormhole definitions confronted,
	Class.Quantum Grav. {\bf 35}, 55009 (2018); 1707.01078.
		
\bibitem{maeda09}		
	Hideki Maeda, Tomohiro Harada and B.J. Carr,	
	Cosmological wormholes,	
	Phys. Rev. D {\bf 79}, 044034 (2009); arXiv: 0901.1153.
			
\bibitem{lemos03}
	J.P.S. Lemos, F.S.N. Lobo and S.Q. de Oliveira, 
	Morris-Thorne wormholes with a cosmological constant, 
	Phys. Rev. D {\bf 68}, 064004 (2003); gr-qc/0302049.	

\bibitem{cataldo08}
	Mauricio Cataldo and Sergio del Campo,
	Two-fluid evolving Lorentzian wormholes,
	Phys. Rev. D {\bf 85},104010 (2012); arXiv: 1204.0753.
	
\bibitem{pha1}
	K.A. Bronnikov and J.C. Fabris, 
	Regular phantom black holes,
	Phys. Rev. Lett. {\bf 96}, 251101 (2006); gr-qc/0511109.

\bibitem{pha2}
	K.A. Bronnikov, H. Dehnen and V.N. Melnikov, 
	Regular \bhs\ and black universes,
	Gen. Rel. Grav. {\bf 39}, 973 (2007); gr-qc/0611022.	
	
\bibitem{pha-em}      
	S.V. Bolokhov, K.A. Bronnikov, and M.V. Skvortsova,
	Magnetic black universes and wormholes with a phantom scalar,
	Class. Quantum Grav. {\bf 29}, 245006 (2012); arXiv: 1208.4619.

\bibitem{pha-sca}   
	K.A. Bronnikov, 
	Scalar fields as sources for wormholes and regular black holes, 
	Particles {\bf 2018}, 1, 56--81 (2018); arXiv: 1802.00098.	

\bibitem{kar94}
	Sayan Kar, Evolving wormholes and the weak energy condition,
	Phys. Rev. D {\bf 49}, 862 (1994). 

\bibitem{sw_kim96}
	Sung-Won Kim, Cosmological model with a traversable wormhole,
	Phys. Rev. D {\bf 53}, 6889 (1996). 

\bibitem{anchor97}
	Luis A. Anchordoqui, Diego F. Torres, Marta L. Trobo, and 
	Santiago E. Perez Bergliaffa, Evolving wormhole geometries,
	Phys. Rev. D {\bf 57}, 829 (1998); gr-qc/9710026.
	
\bibitem{sushkov07}
	Sergey V. Sushkov and Yuan-Zhong Zhang,	
	Scalar wormholes in cosmological setting and their instability,
	Phys. Rev. D {\bf 77}, 024042 (2008); arXiv: 0712.1727.

\bibitem{mokeeva12}	
	Anna Mokeeva and Vladimir Popov,	
	Nonsingular Chaplygin gas cosmologies	in universes connected by a wormhole,
	Grav. Cosmol. {\bf 19}, 57 (2013); arXiv: 1205.1542.	
	
\bibitem{arellano06}	
	A.V.B. Arellano and F.S.N. Lobo, 
	Evolving wormhole geometries within nonlinear electrodynamics, 
	Class. Quantum Grav. {\bf 23}, 5811 (2006); gr-qc/0608003.
	
\bibitem{NED18}	
	K. A. Bronnikov, 
	Nonlinear electrodynamics, regular black holes and wormholes,
	Int. J. Mod. Phys. D {\bf 27}, 1841005  (2018); arXiv: 1711.00087.

\bibitem{fara10}		
	I. Bochicchio and 	Valerio Faraoni,
	A Lema\^{\i}tre--Tolman--Bondi cosmological wormhole,
	Phys. Rev. D {\bf 82}, 044040 (2010); arXiv: 1007.5427.
	
\bibitem{markov} 	
	M.A. Markov and V.P. Frolov,
	Metrics of the closed Friedman world perturbed by electric charge (to the theory of
	 electromagnetic *friedmons*), Teor. Mat. Fiz. {\bf 3}, 3--17 (1970).

\bibitem{bailyn} 
	M. Bailyn, Oscillatory behavior of charge-matter fluids with $e/m > G^{1/2}$,
	Phys. Rev. D {\bf 8}, 1036 (1973)

\bibitem{vickers} 
	P.A. Vickers, Charged dust spheres in general relativity, 
	Ann. Inst. Henri Poincar\'e A {\bf 18}, 137 (1973).

\bibitem{lapch} 
	D.D. Ivanenko, V.G. Krechet, and V.G. Lapchinskii, 
	The dynamics of charged dust in the general theory of relativity,
	Sov. Phys. J. {\bf 16}, 1675--1679 (1973); https://doi.org/10.1007/BF00893659.

\bibitem{khlest}
	Yu.A. Khlestkov, Three types of solutions of the Einstein-Maxwell equations,
	J. Exp. Teor. Fis. {\bf 41} (2), 188 (1975).

\bibitem{shik}
	I.S. Shikin,  An investigation of a class of gravitational fields for a charged dustlike medium,
	J. Exp. Teor. Fis. {\bf 40} (2), 215 (1975).
	
\bibitem{LL} 	
	L.D. Landau and E.M. Lifshitz, {\it The Classical Theory of Fields} 
	(4th ed., Butterworth-Heinemann, 1987).
	
\bibitem{sha08}  
	Alexander Shatskiy, I.D. Novikov and N.S. Kardashev,
	New analytic models of traversable wormholes,
	Phys. Usp. {\bf 51}, 457--464 (2008); arXiv: 0810.0468.
	
\bibitem{sukhan18}
	Yu. A. Khlestkov and L. A. Sukhanova,
	Internal structure of wormholes --- geometric images of charged 
	particles in general relativity, Grav. Cosmol. {\bf 24}, 360 (2018).
				
\bibitem{cle84} 
 	G\'erard Cl\'ement, 
 	A class of wormhole solutions to higher-dimensional general relativity,
 	Gen. Rel. Grav, {\bf 16}, 131 (1984).
 	
\bibitem{bobs} 	
 	K.A. Bronnikov, Block-orthogonal brane systems, black holes and wormholes,  
 	Grav. Cosmol. {\bf 4}, 49 (1998); hep-th/9710207.

\bibitem{dzhu00}
	V. Dzhunushaliev and H.-J. Schmidt,
	Wormholes and flux tubes in the 7D gravity on the principal bundle
	with SU(2) gauge group as the extra dimensions,
	Phys. Rev. D {\bf 62}, 044035 (2000); gr-qc/9911080.

\bibitem{bene03}
	A. DeBenedictis and A. Das.
	Higher-dimensional wormhole geometries with compact dimensions,
	Nucl. Phys. B {\bf 653}, 279 (2003); gr-qc/0207077.

\bibitem{wh-bw}
	K.A. Bronnikov and Sung-Won Kim. 
	Possible wormholes in a brane world,
	Phys. Rev. D {\bf 67}, 064027 (2003).

\bibitem{lobo07}
	Francisco S.N. Lobo. General class of braneworld wormholes,
	Phys. Rev. D {\bf 75}, 064027 (2007).

\bibitem{dzhu14}
	Vladimir Dzhunushaliev and Vladimir Folomeev,
	Kaluza-Klein wormholes with the compactified fifth dimension,
	Mod. Phys. Lett. A {\bf 29}, 1450025 (2014).

\bibitem{dotti14}
	Gustavo Dotti, Julio Oliva, and Ricardo Troncoso.
	Static wormhole solution for higher-dimensional gravity in vacuum,
	Phys. Rev. D {\bf 75}, 024002 (2007).

\bibitem{kuh18}
	Peter K. F. Kuhfittig,
	Traversable wormholes sustained by an extra spatial dimension,
	Phys. Rev. D {\bf 98}, 064041 (2018); arXiv: 1809.01993. 
 
\bibitem{baruah19}
	Anshuman Baruah and Atri Deshamukhya,
	Traversable wormholes in higher-dimensional theories of gravity,
	 J. Phys.: Conf. Ser. {\bf 1330}, 012001 (2019); arXiv: 1904.04928. 

\bibitem{k95}	
	K.A. Bronnikov. Spherically symmetric solutions in D-dimensional dilaton gravity,
	Grav. Cosmol. {\bf 1}, 67 (1995); gr-qc/9505020.

\bibitem{bim97}	
	K.A. Bronnikov, V.D. Ivashchuk and V.N. Melnikov, 
     The Reissner-Nordstr\"om problem for intersecting electric and magnetic
     $p$-branes, Grav. Cosmol. {\bf 3}, 203 (1997); gr-qc/9710054.

\bibitem{k73}
	K.A. Bronnikov, Scalar-tensor theory and scalar charge.
	Acta Phys. Polon. B {\bf 4}, 251 (1973).

\bibitem{hell}
	H. Ellis, Ether flow through a drainhole: a particle model in general relativity,
	J. Math. Phys. {\bf 14}, 104 (1973).

\bibitem{azreg20}
	Mustapha Azreg-A\"{\i}nou,
	Rotating cosmological cylindrical wormholes in GR and TEGR 
	sourced by anisotropic fluids,
	Physics of the Dark Universe {\bf 32}, 100802 (2021); arXiv: 2012.03431.
	
\bibitem{dz-tor}	
	Vladimir Dzhunushaliev, Vladimir Folomeev, Burkhard Kleihaus and Jutta Kunz,
	Thin-shell toroidal wormhole,
	Phys. Rev. D {\bf 99}, 044031 (2019); arXiv:1901.07545.

\bibitem{roman93}	
	T. A. Roman, Inflating Lorentzian wormholes,
	Phys. Rev. D {\bf 47}, 1370 (1993); gr-qc/9211012.	

\bibitem{kard06}	
	N.S. Kardashev, I.D. Novikov and A.A. Shatskiy,
	Astrophysics of wormholes,
	Int. J. Mod. Phys. D {\bf 16}, 909 (2007); astro-ph/0610441.
	
\bibitem{savel15}	
	A.A. Kirillov and E.P. Savelova,	
	Cosmological wormholes,
	Int. J. Mod. Phys.D {\bf 25},1650075 (2016); arXiv: 1512.01450.

\bibitem{extra21}
	K.A. Bronnikov and S.G. Rubin, 
	Local regions with expanding extra dimensions,
	arXiv: 2107.13893.

\end{thebibliography}
\end{document}